\documentclass[fleqn,usenatbib]{mnras}
\usepackage{ifthen,everypage}
\usepackage{graphicx}
\usepackage{amsmath}
\usepackage{multirow}
\usepackage{subcaption}

\title{Radio multifrequency observations of Abell~781 with the WSRT}

\author[B Hugo et al.]{B.~Hugo$^{2,1}$\thanks{E-mail: bhugo@sarao.ac.za},
G.~Bernardi$^{3,1,2}$,
O.M.~Smirnov$^{1,2}$,
D.~Dallacasa$^{4,3}$,
T.~Venturi$^{3}$, \newauthor
M.~Murgia$^{5}$ and R.~F.~Pizzo$^{6}$
\\
$^{1}$Department of Physics \& Electronics, Artillery Road, Rhodes University, Grahamstown, South Africa\\
$^{2}$South African Radio Astronomy Observatory, FIR street, Observatory, Cape Town, South Africa\\
$^{3}$INAF-Istituto di Radioastronomia, via Gobetti 101, 40129, Bologna, Italy\\
$^{4}$Dipartimento di Fisica e Astronomia, Università di Bologna, via P. Gobetti 93/2, I-40129 Bologna, Italy\\
$^{5}$INAF - Osservatorio Astronomico di Cagliari Via della Scienza 5, I-09047 Selargius (CA), Italy\\
$^{6}$ASTRON, the Netherlands Institute for Radio Astronomy, Postbus 2, NL-7990 AA Dwingeloo, The Netherlands
}

\date{This is a pre-copyedited, author-produced PDF of an article accepted for publication in Monthly Notices of the Royal Astronomical Society, Published by Oxford University Press on behalf of the Royal Astronomical Society, following peer review. The version of record has not yet been made available at the time of submission onto the ArXiv.}

\pubyear{2018}

\begin{document}
\label{firstpage}
\pagerange{\pageref{firstpage}--\pageref{lastpage}}
\maketitle

\begin{abstract}

{ The `Main' galaxy cluster in the Abell 781 system is undergoing a significant merger and accretion process with peripheral emission to the north and southeastern flanks of the merging structure}. Here we present a full polarimetric study of this field, using radio interferometric data taken at 21 and 92~cm with the Westerbork Synthesis Radio Telescope, to a sensitivity better than any { 21~cm (\textit{L}-band)} observation to date. {We detect evidence of extended low-level emission of 1.9~mJy associated with the Main cluster at 21~cm, although this detection necessitates further follow-up by modern instruments due to the limited resolution of the Westerbork Synthesis Radio Telescope.} Our polarimetric study indicates that, most likely, the peripheral emission associated with this cluster is not a radio relic.

\end{abstract}


\section{Introduction} 
{Galaxy clusters are some of the largest-scale structures in the Universe, typically spanning a few Mpc. They have masses ranging from $\sim 10^{14}$ up to $\sim 10^{15}$~M$_{\sun}$ \cite[and references therein]{van2019diffuse}, of which only $\sim 3-5\%$ can be associated with luminous matter in constituent galaxies, while $\sim 15-17\%$ in the form of hot ionized gas is detectable through thermal Bremsstrahlung emission in the X-ray regime. The majority ($\sim80\%$) takes the form of dark matter \cite[and references therein]{feretti2012clusters}.}

The shape and curvature in the jets and lobes of Active Galactic Nuclei (AGN) found among galaxy members can be used to infer the motion of constituent galaxies within the cluster, while the study of the Intra-Cluster Medium (ICM) provides insight into the large-scale magnetic fields and physical forces at play during mergers. It is typical to find constituent AGN displaying head-tail, wide- and narrow-angle tail jet morphology. Enabled by multi-wavelength observations, our understanding of cluster evolution has increased dramatically in the past few decades. For instance, radio observations have shown that often there is a significant non-thermal diffuse emission component in merging cluster systems that are sufficiently heated, \citep[therefore detectable in X-rays with integrated energy releases of $10^{63}$ to $10^{64} \text{erg}$,][]{venturi2011elusive}. {Such emission has a very low surface brightness between $\sim 1$ to 0.1~$\mu$Jy~arcsec$^{-2}$ at 1.4~GHz \citep{feretti2012clusters}, and takes the form of broad diffuse emission on scales spanning $100\text{s kpc}$ up to $\sim 1-2$~Mpc \citep[and references therein]{feretti2012clusters}.} It also has a strong morphological correspondence to the emission detectable in X-rays: round in shape and roughly centred at the peak of X-ray luminosity. These are referred to as \textit{radio haloes}. Such radio haloes are detected in roughly 30\% of clusters with integrated X-ray luminosity of $L_x > 5\times10^{44}~\text{erg}~\text{s}^{-1}$ \citep{feretti2012clusters} and are mostly associated to clusters with merger activity. Smaller \textit{mini haloes} can also be found in the less energetic environments of cool-core clusters, { closely related to the core region} of such clusters and typically have sizes less than 0.5 Mpc \cite[and references therein]{van2019diffuse}.

The existence of radio emission on such scales is puzzling. The integrated spectra\footnote{Throughout this paper it is assumed that flux density follows a power law of the form $S(\nu) \propto \nu^{\alpha}$} of radio haloes are in the range $\alpha=-1.2$ to $-1.7$ in the $0.3$ to $1.4$~GHz range \citep{feretti2012clusters}\footnote{We note the spectral index convention we follow is negated with respect to \citet{feretti2012clusters}}. Current estimates on the radiative lifetime of relativistic electrons due to synchrotron and { Inverse Compton (IC)} energy losses are on the order of $10^8$ years at most \citep{sarazin1999energy}. { This is roughly up to 2 orders of magnitude lower than the expected electron diffusion time, assuming an electron diffusion velocity of $\approx 100~\text{km}~\text{s}^{-1}$ \citep{feretti2012clusters}.} { The prevailing theory to their origin suggests the presence of local re-acceleration mechanisms within the ICM through both first and second-order Fermi processes}. First-order processes refer to shock acceleration created in disturbed cluster environments, driving diffuse particle scatter from heterogeneous magnetic fields in both the shock upstream and downstream regions. In contrast, secord-order processes refer to energy gains from turbulence in the ICM. The physical extent over which the diffuse emission is located also precludes that their origin is based in individual galaxy processes. Detailed gamma-ray studies of the Coma cluster suggest that hadronic interactions with {Cosmic Ray (CR) protons in the ICM are not the main origin} of such diffuse emission --- at least not in the case of the giant haloes seen in strong merging clusters. In general, the radio emission from radio haloes does not show significant polarization.

Radio haloes are not the only large-scale diffuse emission that can be associated with merging clusters. \textit{Radio relics} are diffuse sources often seen on the outskirts of clusters (again in a merging state), and they can be found more than $\sim 1~\text{Mpc}$ away from cluster centres \citep{feretti2012clusters}. In general, radio relics have elongated morphologies, and they {can themselves be $100\text{s kpc}$ to $>1\text{Mpc}$ in size \citep{van2019diffuse}.} These sources do not have any direct optical or emitting X-ray counterparts but are often discovered in discontinuities in the X-ray brightness \citep{feretti2012clusters}. They provide perhaps the best evidence for the presence of relativistic particles and strong magnetic fields in very low-density ICM environments, where X-ray sensitivity often precludes a detailed direct study of thermal gas dynamics.

{ Relics provide evidence of radiative ageing}; their radio spectra show clear steepening in the direction of the cluster centre. An excellent example is the 2~Mpc elongated relic on the outskirts of CIZA J2242.8+5301, ranging from $\alpha\approx -0.6$ to $-2.0$. There is also a high degree of polarization across the relic ($50-60\%$) with magnetic field vectors aligned with the relic edge, which is evidence of a well-ordered magnetic field \citep{van2010particle}. This suggests that relics may be driven by shocks and turbulence from merger events, where the shock front compresses the ICM, ordering/amplifying the magnetic field and accelerating relativistic particles. {These elongated relics typically have integrated spectra in the range $\alpha=-1.0$ to $-1.6$ due to low Mach numbers; in agreement with the relic shock model \citep{feretti2012clusters}}.

{The reader is referred to \cite{van2019diffuse} and \cite{feretti2012clusters} for reviews on the topic.}
\section{The curious case of A781} \label{sec:a781intro}
\begin{figure}
\centering
\includegraphics[width=0.5\textwidth]{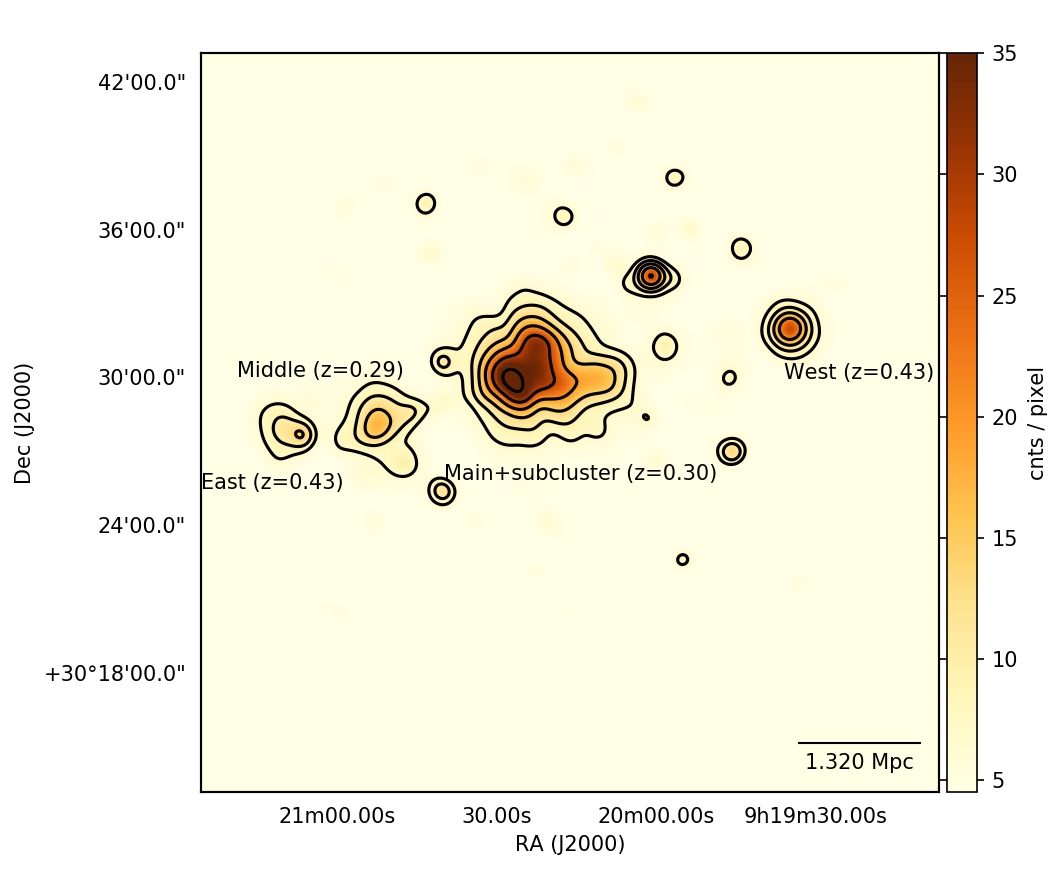}
\caption{\textit{XMM-Newton} MOS1+MOS2 X-ray image. Observation id 0401170101. Here the same labels for the 4 clusters and the merging clusters are used as in previous literature \citep[e.g.,][]{govoni2011large}. The image was convolved with a normalized circular Gaussian of $\sigma=8~\text{arcsec}$ (2 Skypixels). The contours starts from 7~cts~Skypixel$^{-1}$ in steps of a factor of $\sqrt{2}$. { Scale bar drawn the cluster redshift of $z=0.3004$ of the `Main' cluster --- the cluster of interest in this work.}}
\label{fig_xray_overview}
\end{figure}

Abell 781 consists of 4 clusters visible in the X-rays, at least one of them shows clear merger activity. The `Middle' and `Main' clusters, as indicated in Fig.~\ref{fig_xray_overview}, represent an interacting pair, the former showing signs of interaction with smaller structures. Conversely, the `East' and `West' clusters are at different redshifts and might themselves be a pair of objects in a possible long-range interaction. 

In this work we are focusing on the { merging `Main' cluster system} (see Fig.~\ref{fig:final_21_image}). The mass of the Main cluster is $M_{500} = (6.1 \pm 0.5)\times10^{14} M_{\odot}$ \citep{ade2016planck}. Various previous works targeted this Main cluster in Abell 781 in the radio and X-ray. The compact cluster AGN are well studied. Arcsecond scale images at 1.4~GHz of the brightest radio galaxies are presented in \cite{govoni2011large}.

There is a diffuse source in the southeastern part of the Main cluster, as well as two more diffuse sources whose origin is not yet well established. The former has two possible interpretations, namely being either a relic source \citep[e.g.,][]{venturi2011elusive} or a head-tail radio source \citep{botteon2019spectacular}. 

The presence of a giant radio halo has been reported by \cite{govoni2011large} in their analysis of low-resolution JVLA data at 1.4~GHz. However, deep LOw Frequency ARray (LOFAR) and (upgraded) Giant Metrewave Radio Telescope (uGMRT) studies at lower frequencies presented by \cite{botteon2019spectacular} and \cite{venturi2011elusive} point to the contrary. \cite{botteon2019spectacular} place a 50~mJy upper bound on the halo flux density at 143~MHz, while \cite{venturi2011elusive} place an upper bound of $S_{325\text{ MHz}} < 40 \text{mJy}$. These bounds indicate that Abell 781 is an example of one of the high-mass disturbed cluster environments that lack extended radio halo emission --- at least when compared to typical haloes discussed in the literature.

The X-ray luminosity of the cluster is $L_{0.1-2.4\text{kev}}=1.722\times10^{45}\text{ erg }\text{s}^{-1}$ \citep{ebeling1998}. A detailed study of the X-ray emission and its discontinuities is presented by \cite{botteon2019spectacular}. {The study suggests that the Main A781 cluster is undergoing a merger between three smaller clumps}; two in the north-south axis and one responsible for the western bulge in the hot X-ray emission. Their analysis also shows strong evidence of cold fronts at both the south and north edges of the hot X-ray emission. The presence of shock-driven re-acceleration of electrons is still up for debate: the previous analysis only found evidence for a weak shock with a Mach number of $\mathcal{M} < 1.4$ \citep{botteon2019spectacular}.

In this paper, we present sensitive observations carried out with the Westerbork Synthesis Radio Telescope (WSRT) at 21 and 92~cm targeting Abell~781, aiming at characterizing the radio emission of the cluster and its members.

{
Throughout this paper, we will assume $\Lambda$CDM cosmology with $H_0=69.6\text{km }\text{s}^{-1}\text{Mpc}^{-1}$, $\Omega_m = 0.286$, $\Omega_\Lambda = 0.714$. At the cluster redshift of $z=0.3004$, 1~arcsec corresponds to 4.5 kpc. At this redshift, the radio luminosity distance, $D_\text{L}$, is $\sim1570~\text{Mpc}$. 
}
\section{Observation and data reduction} \label{sec:obsdetail}
\subsection{WSRT 21 and 92~cm data}

The Westerbork Synthesis Radio Telescope is an east-west interferometer consisting of 10 fixed-position antennas and 4 re-configurable antennas, each 25~m in diameter, prime focus and equatorial-mounted. The 10 fixed-position antennas are regularly separated by 144m, with a minimum distance of 36m between the last fixed antenna and the first re-configurable antenna. The maximum spacing is 2.7 km. 

The field was observed prior to wide-field phased-array receiver upgrades \citep[see e.g.][]{verheijen2008apertif} and used the old 21 cm (1321--1460~MHz) and 92 cm (320--381~MHz) receivers. The 21 and 92 cm correlators have frequency resolutions of 312.50 and 78.125 kHz respectively. Both 21 cm and 92 cm receivers employ a linearly-polarized feed system. Observation details are summarized in Table~\ref{tab_obsinfo}. 
\begin{table*}
  \centering
  \begin{tabular}{ccccccc}
    \hline \\
    Band & Config & Obs ID & Target & Span (UTC) & J2000 RA & J2000 DECL\\
    \hline \\
    92cm & 36m & 11200314 & 3C147 & 2012 Jan 17 18:30:50 -- 18:45:50 & $05^\text{h}42^\text{m}36.135^\text{s}$ & $+49^\circ 51'07"$\\
    & & 11200315 & DA240 & 2012 Jan 17 18:49:30 -- 19:04:30 & $07^\text{h}49^\text{m}48.017^\text{s}$ & $+55^\circ 54'22"$\\
    & & 11200316 & A781 & 2012 Jan 17 19:07:50 -- Jan 18 07:06:50 & $09^\text{h}20^\text{m}25.401^\text{s}$ & $+30^\circ 30'07"$\\
    & & 11200317 & 3C295 & 2012 Jan 18 07:12:40 -- 07:27:40 & $14^\text{h}11^\text{m}20.652^\text{s}$ & $+52^\circ 12'09"$\\
    & 48m & 11200407 & 3C147 & 2012 Jan 23 18:07:10 -- 18:22:10 \\
    & & 11200408 & DA240 & 2012 Jan 23 18:25:50 -- 18:40:50 \\
    & & 11200409 & A781 & 2012 Jan 23 18:44:10 -- Jan 24 06:43:10 \\
    & & 11200410 & 3C295 & 2012 Jan 24 06:49:00 -- 07:04:00 \\
    & 60m & 11200434 & 3C147 & 2012 Jan 24 18:03:20 -- 18:18:20 \\
    & & 11200435 & DA240 & 2012 Jan 24  18:22:00 -- 18:37:00 \\
    & & 11200436 & A781 & 2012 Jan 24 18:40:20 -- Jan 25 04:22:00 \\
    & & 11200439 & A781 & 2012 Jan 25 06:05:40 -- 06:39:20 \\
    & & 11200440 & 3C295 & 2012 Jan 25 06:45:10 -- 07:00:10 \\
    & 96m & 11201063 & 3C147 & 2012 Feb 13 16:44:40 -- 16:59:40 \\
    & & 11201064 & DA240 & 2012 Feb 13 17:03:20 -- 17:18:20 \\
    & & 11201065 & A781 & 2012 Feb 13 17:21:40 -- Feb 14 05:20:40 \\
    & & 11201066 & 3C295 & 2012 Feb 14 05:26:30 -- 05:41:30 \\
    & 84m & 11201079 & 3C147 & 2012 Feb 14 16:40:40 -- 16:55:40 \\
    & & 11201080 & DA240 & 2012 Feb 14 16:59:20 -- 17:14:20 \\
    & & 11201081 & A781 & 2012 Feb 14 17:17:40 -- Feb 15 05:16:40 \\
    & & 11201082 & 3C295 & 2012 Feb 15 05:22:30 -- 05:37:30 \\
    & 72m & 11202096 & 3C147 & 2012 Mar 31 13:39:50 -- 13:54:50 \\
    & & 11202097 & DA240 & 2012 Mar 31 13:58:30 -- 14:13:30 \\
    & & 11202098 & A781 & 2012 Mar 31 14:16:50 -- Apr 01 02:15:50 \\
    & & 11202099 & 3C295 & 2012 Apr 01 02:21:40 -- 02:36:40 \\
    21cm & 36m & 11200302 & 3C48 & 2012 Jan 16 18:48:30 -- 19:03:30 & $01^\text{h}37^\text{m}41.30^\text{s}$ & $+33^\circ 09'35"$\\
    & & 11200303 & A781 & 2012 Jan 16 19:11:40 -- 23:56:10 \\
    & & 11200305 & 3C286 & 2012 Jan 17 07:37:20 -- 07:52:20 & $13^\text{h}31^\text{m}08.29^\text{s}$ & $+30^\circ 30'33"$\\
    & 54m & 11202012 & 3C48 & 2012 Mar 28 14:05:20 -- 14:20:20 \\
    & & 11202013 & A781 & 2012 Mar 28 14:28:40 -- Mar 29 02:27:40 \\
    & & 11202014 & 3C286 & 2012 Mar 29 02:33:00 -- 02:48:00 \\
    & 72m & 11202119 & 3C48 & 2012 Apr 01 13:49:40 -- 14:04:40 \\
    & & 11202120 & A781 & 2012 Apr 01 14:12:50 -- Apr 02 02:11:50 \\
    & & 11202121 & 3C286 & 2012 Apr 02 02:17:10 -- 02:32:10 \\
    \hline
  \end{tabular}
  \caption{Summary of relevant observations of A781 and celestial calibrator fields taken with WSRT in various configurations at 21 and 92 cm in 2012.}
  \label{tab_obsinfo}
\end{table*}

The Westerbork Synthesis Radio Telescope uses a programmable temperature-stabilized noise diode to correct for the time-variable electronic gains. See for instance \cite{casse1974synthesis} and \cite{bos1981digital} for a brief description. The frequency-dependent response of the system is calibrated with a strong celestial source. Delays and phases on crosshands are corrected with a strongly polarized celestial source before leakages are corrected using an unpolarized source. The first-order on-axis linear feed calibration strategy we followed is discussed in more detail in \cite{hales2017calibration}. Parallactic angle corrections are not required, because the equatorial mounts of the WSRT imply that the sky does not rotate with respect to the receiver as a function of the hour angle. This further implies that we have to rely on polarized sources with known polarization angles to calibrate crosshand phases and correct for the system-induced ellipticity and its interplay with the linear polarization angle. 
%

The 21~cm band data are calibrated for the complex bandpass response of the system using 3C48, which has limited linear polarization of 0.5~Jy, assuming the following model \citep{perley2013accurate}:

\begin{equation*}
\log{S} = 1.3324 -0.7690\log{\nu_\text{G}} -0.1950\log^2{\nu_\text{G}} +0.059\log^3{\nu_\text{G}}.
\end{equation*}
Here $\nu_\text{G}$ is given in GHz and the flux density, $S$, in Jy. The system ellipticity (crosshand phase) calibration is performed using the strongly linearly polarized source 3C286. The source is assumed to have a constant polarization angle of $\approx 33^\circ$ across the passband. First-order leakages are corrected using 3C48. We estimate that after correction the total quadrature sum of Stokes Q, U and V to I of this marginally polarized source ranges between 0.025~\% and 0.006~\% across the passband.

The 92~cm data are calibrated for the complex bandpass using the unpolarized source 3C147, assuming the frequency response \citep{perley2013accurate}: 

\begin{equation*}
\log{S} = 1.4616 -0.7187\log{\nu_\text{G}} -0.2424\log^2{\nu_\text{G}} +0.079\log^3{\nu_\text{G}}.
\end{equation*}
DA240 is known to be highly polarized --- parts of its western lobe are above 60\% polarized, see \cite{tsienda240}. However, the source can be resolved by the WSRT at 92~cm. Subsequently, only leakages (off-diagonal terms) can be corrected, for which the unpolarized (at 92~cm) 3C138 is used. After correction, we estimate the quadrature leakages to vary between 0.19~\% and -0.05~\% across the passband.

{ Both the 92~cm and 21~cm data reductions were performed using the containerised astronomy workflow management framework \textsc{Stimela} v0.3.1} \footnote{ Available from \url{https://github.com/SpheMakh/stimela}. \textsc{Stimela} is a pipelining framework for radio astronomy which wraps tasks from a wide variety of heterogeneous packages into a common \textsc{Python} \citep{10.5555/1593511} interface. The often-complicated compilation and software dependencies of these packages are isolated through containerisation platforms, such as \textsc{Docker} \citep{merkel2014docker} and \textsc{Singularity} \citep{kurtzer2017singularity}, allowing a heterogenous set of often-conflicting packages to be accessible through a single workflow interface.} \citep{makhathini_phdthesis}. Calibration was performed using the \textsc{Common Astronomy Software Applications (CASA)} v4.7 \citep{mcmullin2007casa}. We first identified and flagged Radio Frequency Interference using the \textsc{AOFlagger} package \citep{offringa2010aoflagger}. 

After applying the complex bandpass, $5^\circ12'$ wide images were generated using the \textsc{WSClean} package \citep{offringa2014wsclean} with uniform weights. Both the 92 and 21 cm data were imaged at the same resolution and size to simplify the derivation of spectral index maps. All the 144~m redundant spacings were included in the synthesis in order to improve sensitivity. To account for the apparent spectral variation over the observation bandwidth due to the antenna primary beam we enabled the multi-frequency deconvolution algorithm \citep{offringa2014wsclean}. We used the imager's auto-thresholding deconvolution (set to $1\sigma$) criterion which stops deconvolution based on the median absolute deviation of the residual map.  A model of the field was derived as a list of fitted Gaussian components brighter than $5~\sigma$ using the \textsc{pyBDSF} source extractor \citep{mohan2015pybdsf}. After predicting model visibilities using \textsc{Meqtrees} \citep{noordam2010meqtrees}, self-calibration was performed by solving for phases with 180~s and 60~s solution intervals for 21~cm and 92~cm data respectively. The datasets were then re-imaged to construct an improved model for a second round of phase calibration, again fitting Gaussians above $5\sigma$ and using solution intervals of 30~s and 10~s respectively.

\begin{figure*}
\centering
\includegraphics[width=.75\textwidth]{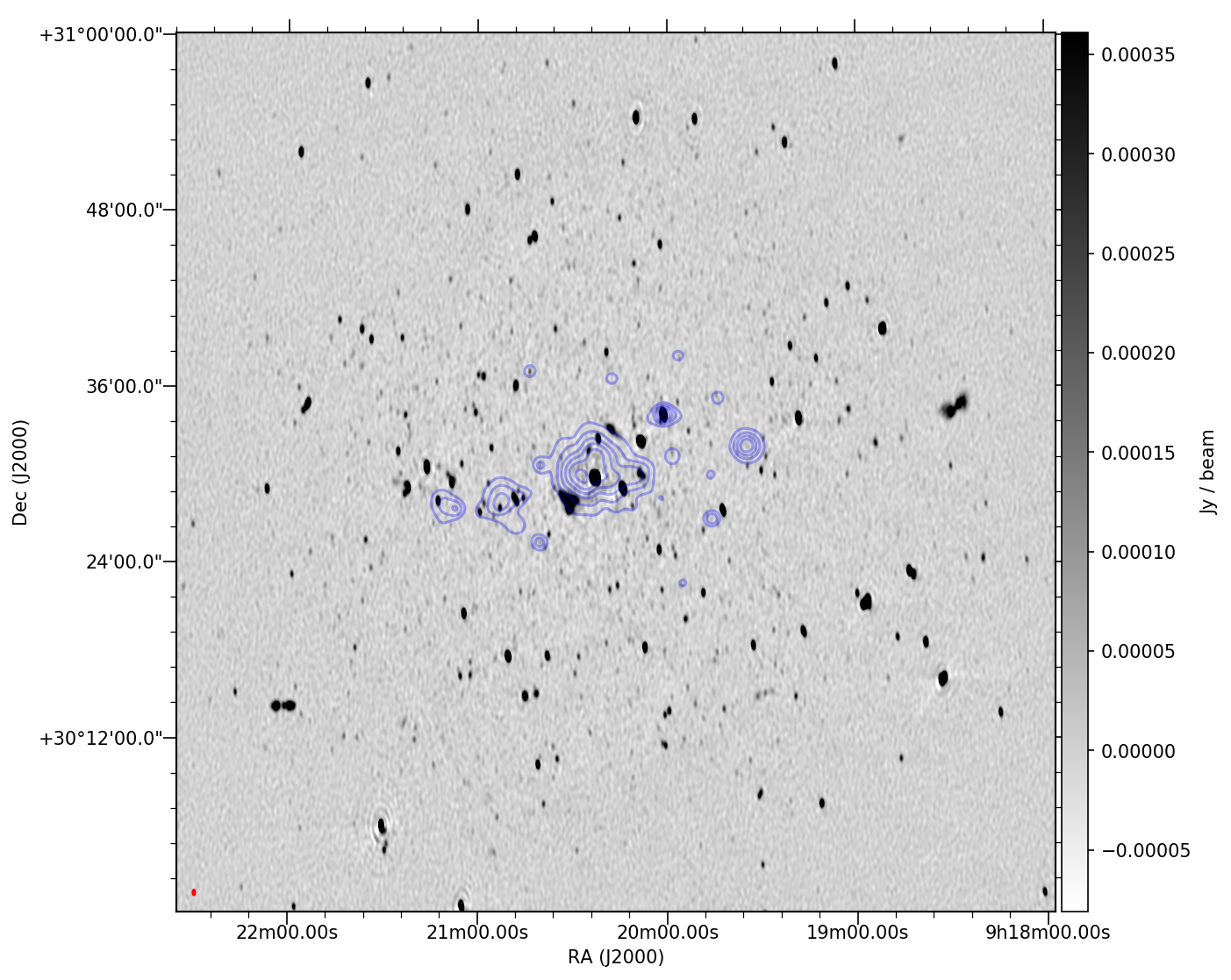}
\includegraphics[width=.75\textwidth]{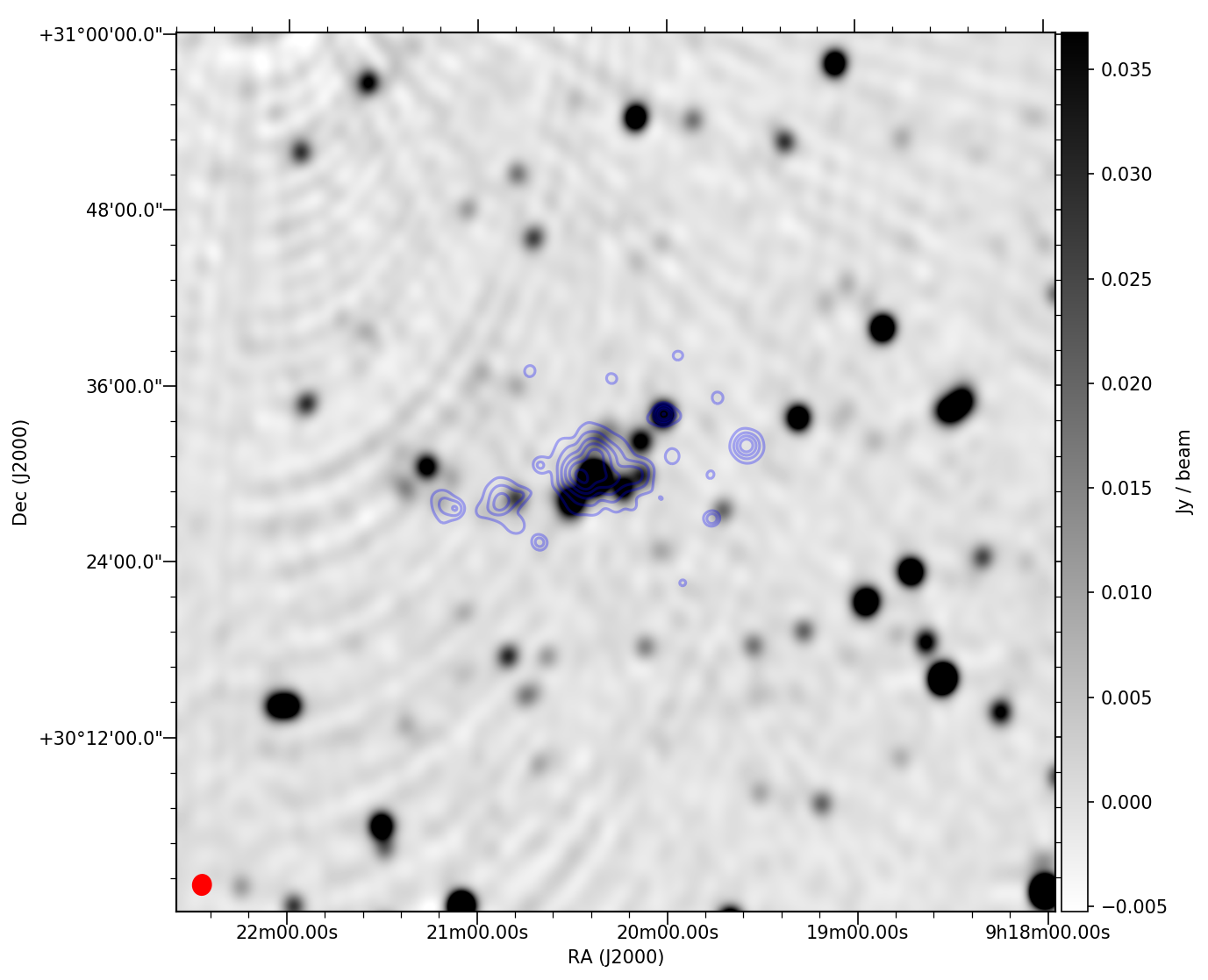}
\caption[Overview radio images 21 and 92~cm]{$1^\circ$-wide 21~cm (top) and 92~cm (bottom) images cropped and centred on the cluster. The 21~cm synthesized beam is $23.2\times10.4~\text{arcsec}$ and the noise rms 12~$\mu$Jy~beam$^{-1}$. The 92~cm synthesized beam is $83\times74~\text{arcsec}$  and the noise rms is 1.5~mJy~beam$^{-1}$ (calculated in an empty patch where the main contribution are direction-dependent calibration artefacts from a bright source to the top left of the zoomed-in region). Both images shown here are not corrected for primary beam attenuation. Since the 92~cm data is only useful to estimate a spectral index map and not used for the detection of faint diffuse emission we did not perform further direction-dependent calibration. { We over plot the contours derived for fig.~\ref{fig_xray_overview} in faint blue for reference --- the \textit{XMM-Newton} exposure used only extends over the central region of the radio map.}}
\label{fig:final_21_image}
\end{figure*}

Finally, residual amplitude and phase self-calibration was performed using a 7~minute solution interval. The final 21~cm and 92~cm uniform-weighted images are shown in Fig.~\ref{fig:final_21_image}.

The 92~cm synthesized beam is too large, even at uniform weighting, to accurately model and subtract bright AGN cluster members at this frequency. The 92~cm data is, however, useful in estimating the spectral profiles of the compact emission. On the other hand, the resolution of the 21~cm data enables us to produce an image at an intermediate resolution of $25.8\times12.1~\text{arcsec}$ using Briggs \citep{briggs1995high} weights of -0.25 (compact sources brighter than $2\sigma$ were subtracted from the visibilities from modelling at highest possible resolution prior to imaging). This image has better sensitivity to extended structure and was used to assess the presence of diffuse emission. 

The 21~cm and 92~cm Briggs -2.0 maps have synthesized beams of $23.2\times10.4~\text{arcsec}$ and $83\times74~\text{arcsec}$ respectively at uniform weighting\footnote{It is worth noting this is somewhat worse than quoted by the {WEsterbork Northern Sky Survey (WENSS) \citep{1997A&AS..124..259R}}, we applied a circular Gaussian taper to improve the synthesized beam shape. This resulted in a decrease in resolution by roughly a factor of 2}. The area of the synthesized beam is approximately given by an elliptical Gaussian and is $\Omega_{21\text{cm}}=20.58~\text{px}$ and $\Omega_{92\text{cm}}=523.97~\text{px}$ respectively\footnote{The sampling here is kept to 3.64~arcsec --- consistent with the 21cm maps to simplify computing Spectral Index (SPI) maps in the analysis.}, where $\theta_l$ and $\theta_m$ are the fitted \texttt{BMAJ} and \texttt{BMIN} at full width half maximum of the synthesized beam in pixels:

\begin{equation*}
    \Omega_b = \int \theta_b dldm \approx \frac{\pi\theta_l\theta_m}{4\ln{2}}.
\end{equation*}

\subsection{WSRT / NVSS flux scale comparison}
In order to quantify the error on the absolute flux scale of our calibration we cross-match the primary-beam-corrected population of compact sources with that of the { National Radio Astronomy Observatory (NRAO) Very Large Array (JVLA or VLA interchangeably) Sky Survey} \citep[NVSS,][]{condon1998nrao}, integrated to the lower resolution of the NVSS in the case of our 21~cm data. The source catalogue is obtained by running \textsc{pyBDSF} \citep{mohan2015pybdsf} to extract the population above 20 sigma using adaptive thresholding. { The WSRT power beam attenuation was corrected according to the following analytical model, prior to catalogue fitting:}

\begin{equation}
  \label{eqn:wsrtpbeam}
  B(\theta, \nu_\text{G}) = \cos^6{(65.0\nu_\text{G}\theta)}
\end{equation}

{ Here $\nu_\text{G}$ is the frequency in gigahertz and theta the evaluated angular separation from the pointing centre.}

\begin{figure}
    \centering
    \includegraphics[width=0.45\textwidth]{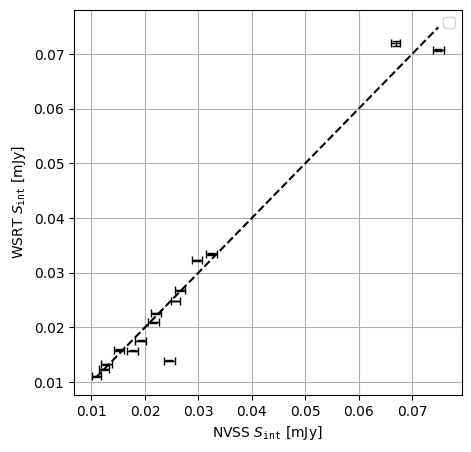}
    \caption{cross-matched fluxes between WSRT 21~cm and the NVSS, integrated to the lower resolution of the NVSS. Error bars indicate $1\sigma$ fit errors taken from the \textsc{pyBDSF} catalog fitting routine. Only sources above $20\sigma$ are shown. The WSRT 21~cm population was scaled to the NVSS frequency assuming a population spectral index of -0.7, typical to AGN at this frequency.}
    \label{fig:nvsswsrt21cmX}
\end{figure}

The flux cross-match is shown in Fig.~\ref{fig:nvsswsrt21cmX}. We obtain a match in flux scale with an average absolute error of 7.13\%. This cross-match error on the absolute flux scale of the NVSS and WSRT 21~cm maps corresponds well to the second measure of absolute flux scale error we estimate by transfer calibration from 3C48 (observed prior to target) onto 3C286 (observed after target). The absolute error in transfer scale to the scale stated by \cite{perley2013accurate} is 6.20\% \footnote{This moderate error could be the result of not having access to a gain calibrator for our observations to monitor for amplitude stability on the system, however out of band linearity limitations are known, for instance, with the MeerKAT system when working at {\textit{L}-band (856--1712~MHz)} which is dominated by Global Navigation Satellite System transmitters. The error quoted here could be a combination of both.}. We similarly quantify the error on the flux scale of the 92~cm data, which was calibrated using 3C147 (observed prior to target), and transferred onto 3C295 (observed after the target). The absolute error to the scale of \cite{perley2013accurate} was found to be 2.39\% on average across the passband. We will assume a 10\% error margin of the \cite{perley2013accurate} scale, which is used throughout as an upper bound to the errors when computing powers and spectral indices. 

\begin{figure}
    \centering
    \includegraphics[width=0.45\textwidth]{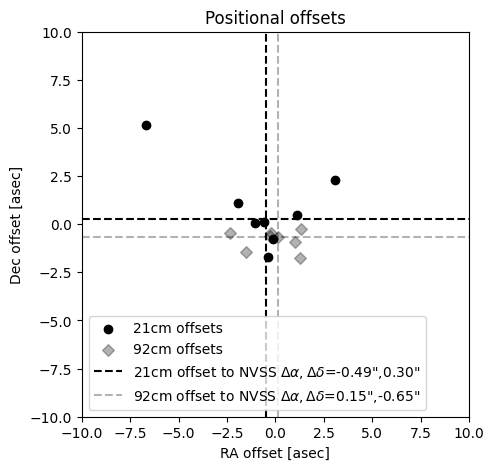}
    \caption{Positional cross-match to the NVSS within the resolution of the NVSS. Here we use the fitted positions and errors given by \textsc{pyBDSF}. In both cases of 21 and 92~cm the positional offset is a fraction of the instrument resolution. The dashed lines indicate the median offset for the selection of compact sources.}
    \label{fig:posxnvss}
\end{figure}

We further compare the positional accuracy by cross-matching to the NVSS. To minimize the positional uncertainties brought about by extended sources we select only compact sources to cross-match. We define a `compactness' criterion by measuring the ratio between integrated flux to peak flux using a \textsc{pyBDSF}-fitted catalogue. A ratio close to unity for a high SNR source indicates that the source is compact. We select such sources that are within $\pm30\%$ of unity to measure positional accuracy, shown in Fig.~\ref{fig:posxnvss}.

\subsection{Archival VLA 21~cm A, C and D configuration data}
The 21~cm WSRT data lacks the necessary resolution to show the structure of the AGNs associated with this cluster. As a possible complementary source of information, we reduced the same archival JVLA \textit{L}-band\footnote{Digitized bandwidth is that for the JVLA prior to updates to its correlator \citep{perley2011expanded}: 1355.5--1447.6~MHz for projects AB699 and AM469 and 1452.4--1527.4~MHz for project AO048 in two disjoint spetral windows.} data used in the analysis by \cite{govoni2011large}.
Details of which are summarized in Table~\ref{tab_obsinfoVLA}. 
\begin{table*}
  \centering
  \begin{tabular}{clccrcc}
    \hline \\
    Conf. & Bandwidth (MHz) & Obs ID & Target & Span (UTC) & J2000 RA & J2000 DECL \\
    \hline \\
    D & 1355.525-1377.4, & AM469 & 1331+305 (3C286) & 1995 Mar 15 05:54:40--05:59:50 & $13^\text{h}31^\text{m}08^\text{s}$ & $+30^\circ 30'32''$ \\
      & 1425.725--1447.6 &       & & 08:00:30--08:05:50 & & \\
      & &       & & 12:35:10--12:41:00 & & \\
      & &       & A0781 & 1995 Mar 15 03:55:10--04:10:20 & $09^\text{h}20^\text{m}23^\text{s}$ & $+30^\circ 31'09''$ \\
      & &       & 0842+185 & 1995 Mar 15 03:03:10--03:06:20 & $08^\text{h}42^\text{m}05^\text{s}$ & $+18^\circ 35'40''$ \\
      & &       &          & 04:12:10--04:13:20 & & \\
    A & 1355.525--1377.4, & AB699 & A0781 & 1994 Apr 20 00:02:00-00:31:40 & $09^\text{h}20^\text{m}23.7^\text{s}$ & $+30^\circ31'09''$ \\
      & 1425.725--1447.6 &       & 1331+305 (3C286) & 1994 Apr 29 01:32:00--01:35:10 & $13^\text{h}31^\text{m}08.2873^\text{s}$ & $+30^\circ30'32.9590''$ \\
      & &       & & 11:37:50--11:42:40 & & \\
      & &       & & 1994 Apr 20 11:14:40--11:18:20 & & \\
      & &       & & 04:58:40--05:01:50 & & \\
      & &       & 1438+621 & 1994 Apr 29 09:40:20--09:41:40 & $14^\text{h}38^\text{m}44.7873^\text{s}$ & $+62^\circ 11'54.397''$\\
      & &       & & 10:42:50--10:44:00 & & \\
      & &       & & 11:15:00--11:16:20 & & \\
      & &       & & 11:35:30--11:36:40 & & \\
      & &       & & 1994 Apr 20 08:24:10--08:25:20 & & \\
      & &       & & 09:27:40--09:28:50 & & \\
    \hline \\
	Config & Bandwidth (MHz) & Obs ID & Target & Span (UTC) & B1950 RA & B1950 DECL\\
    \hline \\
    C & 1452.4--1477.4, & AO048 & 0781AB & 1984 May 05 02:49:00--05:25:30 & $09^\text{h}17^\text{m}23.30^\text{s}$ & $+30^\circ44'05''$ \\
      & 1502.4--1527.4 &        & 1328+307 (3C286) & 1984 May 05 08:32:00--08:35:00 & $13^\text{h}28^\text{m}49.657^\text{s}$ & $+30^\circ 45'58.64''$\\
      & &        & & 08:52:00--08:55:00 & & \\
      & &        & & 09:16:00--09:18:00 & & \\
      & &        & & 11:15:30--11:17:30 & & \\
      & &        & & 11:33:30--11:35:30 & & \\
      & &        & & 11:56:30--11:58:30 & & \\
      & &        & 0851+202 & 1984 May 05 02:41:30--02:43:00 & $08^\text{h}51^\text{m}57.253^\text{s}$ & $+20^\circ 17'58.44''$ \\
      & &        & & 02:59:00--03:02:00 & & \\
      & &        & & 05:06:00--05:15:30 & & \\
      & &        & & 05:31:30--05:33:30 & & \\
      \hline
  \end{tabular}
  \caption{{Archival JVLA observation details. Here only the details of the relevant calibrators and targets field are shown.}}
  \label{tab_obsinfoVLA}
\end{table*}


The data were flagged and calibrated with the \textsc{Common Astronomy Software Applications} (\textsc{CASA}) v4.7 \citep{mcmullin2007casa} through \textsc{Stimela} v0.3.1 \citep{makhathini_phdthesis}. We applied flags to instances of equipment failure and shadowing. Throughout we used 3C286 to set the flux scale of the observation \citep{perley2013accurate} and to calibrate the frequency response of the system. Since the JVLA has circular \textit{L}-band feeds it is not strictly necessary to take a polarization model into account\footnote{In the circular basis the diagonal correlations LL and RR measure $I\pm V$  \citep[e.g.,][]{smirnov2011revisiting} and is insensitive to the linearly polarized flux of the celestial calibrator 3C286}. Although the data is taken at multiple epochs, 3C286 is known to be a very stable calibrator --- within 1\% over the duration of 30 years \citep{perley2013accurate} --- and thus the following model can be assumed for all three datasets:

\begin{equation*}
\log{S} = 1.2515 -0.4605\log{\nu_\text{G}} -0.1715\log^2{\nu_\text{G}} +0.0336\log^3{\nu_\text{G}}
\end{equation*}

The time-variable gain on 3C286 was calibrated by correcting for gains at 30~s intervals, before computing a single normalized bandpass correction for the entire observation. We did not correct the data for polarization leakages, since we are primarily interested in supplementing our analysis with source morphological information. Unlike the WSRT observations, the time-variability of the electronics, especially its phase, needs to be calibrated with a celestial calibrator. We used 0842+185, 1438+621 and 0851+202 for D, A and C configuration data respectively.

Again, we used the multi-frequency deconvolution algorithm implemented in \textsc{WSClean} \citep{offringa2014wsclean}. We enabled widefield corrections at default setting, synthesized a $70.0'$ image and deconvolved to an auto-threshold of $1\sigma$, using Briggs \cite{briggs1995high} weighting with robustness 0.0. The \texttt{CLEAN} residuals have an rms noise of 55~$\mu$Jy~beam$^{-1}$.

\section{Results and discussion}
\label{sec:results}
In this section, we discuss the Main cluster as a whole, a study of the field source polarimetry, { the preliminary detection of a candidate halo} and lack of detection of relic emission.

\begin{figure*}
\centering
\begin{minipage}{0.6\textwidth}
\includegraphics[width=\textwidth]{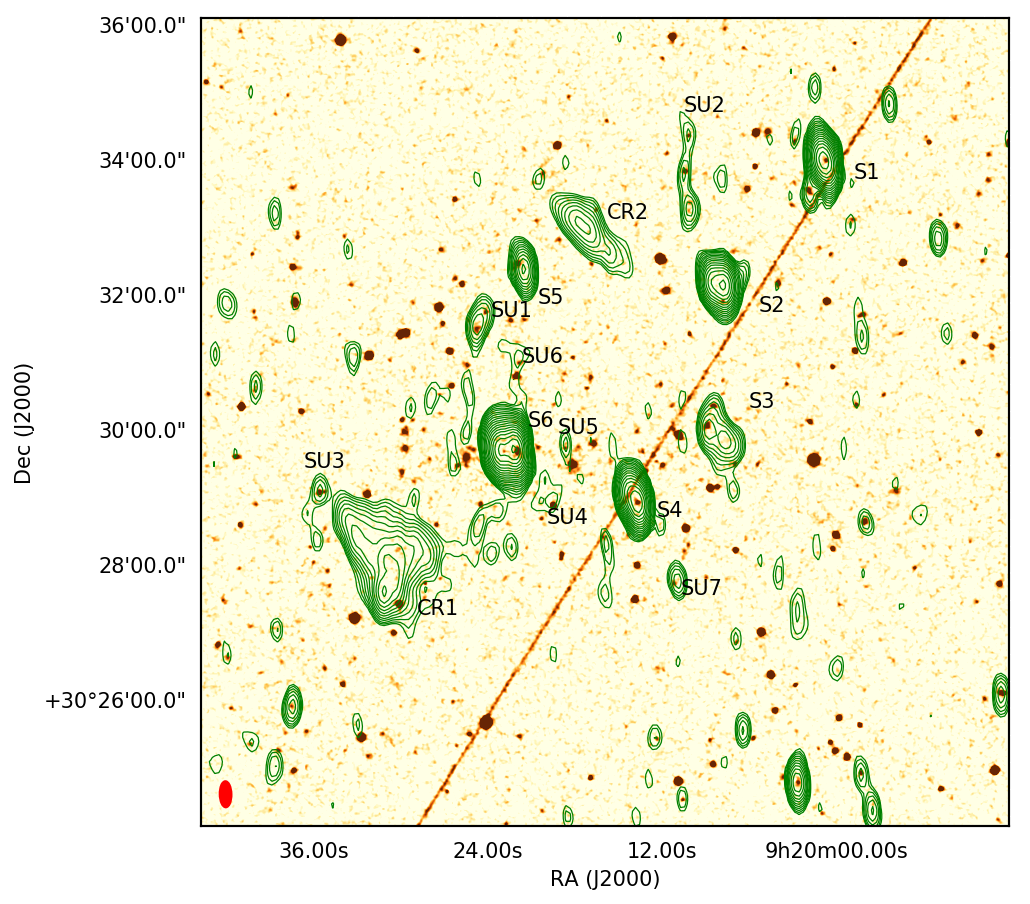} 
\includegraphics[width=\textwidth]{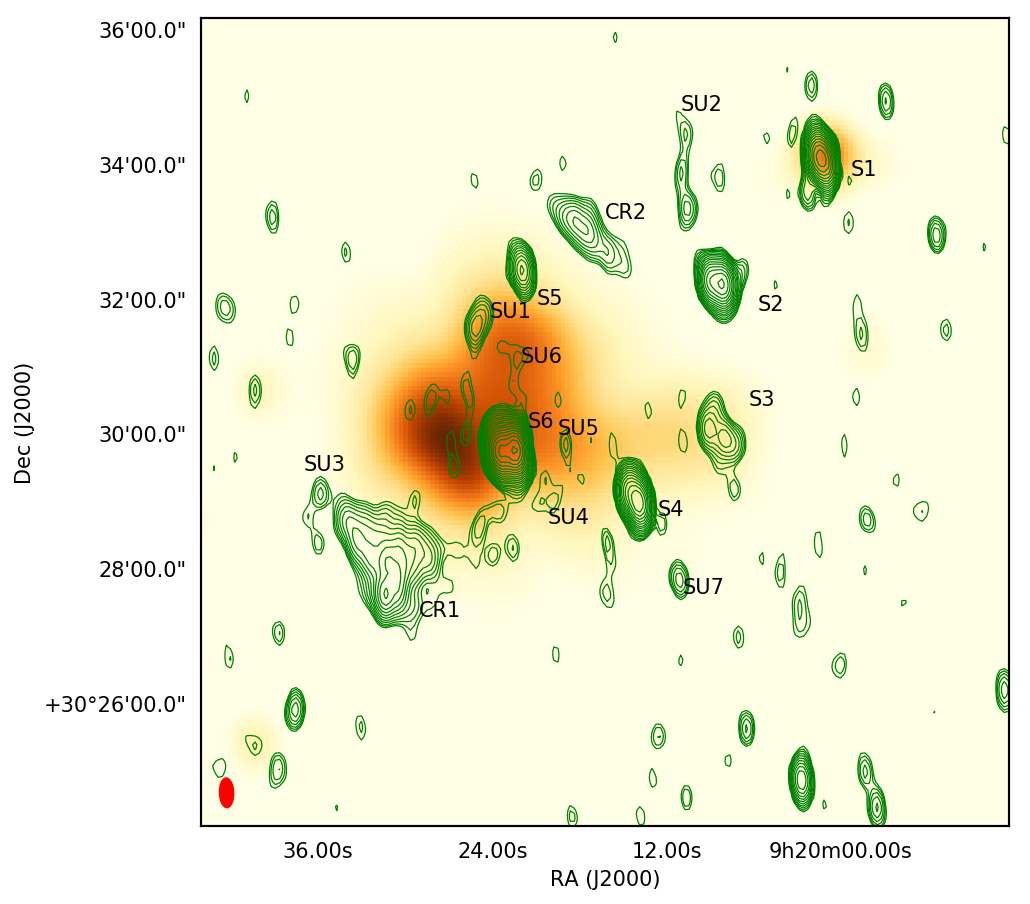}
\end{minipage}%
\begin{minipage}{0.25\textwidth}
\includegraphics[width=\textwidth]{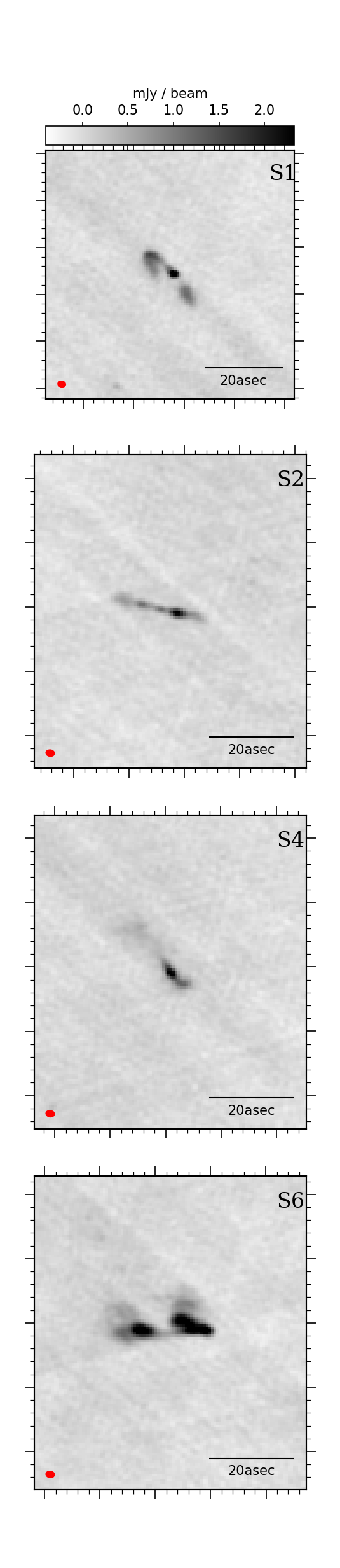}
\end{minipage}
\caption{\textit{Top left}: Radio contours from Fig.~\ref{fig:final_21_image} overlaid over Digitized Sky Survey II POSS2 optical red plate. Contours are drawn starting at 60~$\mu$Jy~beam$^{-1}$ in steps of $\sqrt{2}$. \textit{Bottom left}: Radio contours from Fig.~\ref{fig:final_21_image} are overlaid over \textit{XMM-Newton} X-ray from Fig.~\ref{fig_xray_overview}. \textit{Right}: Higher resolution (synthesized beam of $1.935\times1.454~\text{arcsec}$) tiles of (top to bottom) S1, S2, S4 \& S6, as observed with the JVLA A+C+D configurations between 1.354 and 1.515~GHz, rms noise 54.68~$\mu$Jy~beam$^{-1}$.}
\label{fig:21_uniform_optical}
\end{figure*}

Fig.~\ref{fig:final_21_image} shows the 1.4~GHz image covering the whole WSRT primary beam and the corresponding field at 346~MHz. A zoom into the central field overlaid on the Sloan Digitized Sky Survey 12$^{\rm th}$ release \citep{alam2015eleventh} is shown in Fig.~\ref{fig:21_uniform_optical}. Bright radio sources are labelled and listed, along with their optical counterparts (where available), in Table~\ref{tab_photoredshifts}. \footnote{The radio sources at 1.4~GHz follow the same naming convention used in \cite{venturi2011elusive}}. {Compact sources S1--S6 seen at 325 and 610~MHz by \cite{venturi2011elusive} are visible in our image too. These bright sources (S1--S6) have corresponding optical counterparts, apart from S2. We have labelled those fainter sources with clear corresponding optical counterparts as SU1 through SU7. Sources S3, S4, S5, S6, SU5, SU7, SU2 are at a similar redshift to the cluster within the quoted \citep{beck2016photometric} error bars of the photometric redshifts ($3\sigma=0.06$), as shown in Table~\ref{tab_photoredshifts}. The spectroscopic redshift for S1 is substantially different to the confirmed cluster members and indicates that the source is not part of the Main cluster, but a background AGN. Based on the lack of corresponding optical counterparts and their radio morphology two candidate relics are labelled as CR1 and CR2. Both candidate relics are on the outskirts of the heated cluster medium, while the bright AGN S6 is clearly offset from the peak X-ray emission. The JVLA high-resolution tiles in Fig.~\ref{fig:21_uniform_optical} highlight the extended nature of the AGN S6, which is barely resolved by the WSRT.}


The redshifts of the previously identified radio sources S1, S3, S4, S5 and S6 are listed in Table~\ref{tab_photoredshifts}. S2 does not have any apparent optical counterpart and is, therefore, not included in the table. Sources S3--S6, SU1, SU2, SU5 and SU7 are very likely all cluster members.
\begin{table}
  \centering
  \begin{tabular}{llll}
  \hline \\
  Source ID & RA$_{\rm J2000}$ & DEC$_{\rm J2000}$ & $z$ \\
  \hline \\
  S1 & $9^\text{h}20^\text{m}1.2^\text{s}$ & $+30^\circ 34'5.3"$ & 1.305$^{\rm s}$ \\
  S3 & $9^\text{h}20^\text{m}9.2^\text{s}$ & $+30^\circ 30'8.1"$ & 0.303$^{\rm s}$ \\
  S4 & $9^\text{h}20^\text{m}14^\text{s}$ & $+30^\circ 28'59.4"$ & 0.297$^{\rm s}$ \\
  S5 & $9^\text{h}20^\text{m}22.4^\text{s}$ & $+30^\circ 32'30.7"$ & 0.304$^{\rm s}$ \\
  S6 & $9^\text{h}20^\text{m}22.3^\text{s}$ & $+30^\circ 29'43.3"$ & 0.293$^{\rm s}$ \\
  SU1 & $9^\text{h}20^\text{m}25.2^\text{s}$ & $+30^\circ 31'31.7"$ & 0.304$^{\rm s}$ \\
  $\text{SU2}_{\rm s}$ & $9^\text{h}20^\text{m}10.6^\text{s}$ & $+30^\circ 33'26.0"$ & 0.3(9)$^{\rm p}$ \\ 
  $\text{SU2}_{\rm m}$ & $9^\text{h}20^\text{m}10.9^\text{s}$ & $+30^\circ 33'54.0"$ & 0.2(5)$^{\rm p}$ \\
  $\text{SU2}_{\rm n}$ & $9^\text{h}20^\text{m}10.7^\text{s}$ & $+30^\circ 34'26.1"$ & 0.2(9)$^{\rm p}$ \\
  SU3 & $9^\text{h}20^\text{m}35.9^\text{s}$ & $+30^\circ 29'5.0"$ & 0.1(5)$^{\rm p}$ \\
  SU5 & $9^\text{h}20^\text{m}19.0^\text{s}$ & $+30^\circ 29'47.0"$ & 0.2(8)$^{\rm p}$ \\
  SU6 & $9^\text{h}20^\text{m}22.2^\text{s}$ & $+30^\circ 31'02.3"$ & 0.1(9)$^{\rm p}$ \\
  SU7 & $9^\text{h}20^\text{m}11.4^\text{s}$ & $+30^\circ 27'47.5"$ & 0.2(8)$^{\rm p}$ \\
  \hline
  \end{tabular}
  \caption{Sources S1--S6 were already observed by \citet{venturi2011elusive}. Source SU2 has three corresponding optical sources $\text{SU2}_{\rm s}$, $\text{SU2}_{\rm m}$, $\text{SU2}_{\rm n}$, for the southern, middle and northern source respectively. SU4 had no redshift information and was excluded from the list. Spectroscopic ($^{\rm s}$) and photometric ($^{\rm p}$) redshifts are from the Sloan Digitized Sky Survey 12$^{\rm th}$ release \citep{alam2015eleventh}. { The $3\sigma$ uncertainty level on the photometric redshifts are indicated in parentheses.}}
  \label{tab_photoredshifts}
\end{table}

In studies done to date, the orientation of, and the emission mechanisms behind the CR1 complex are still not clearly established. The complex spans about 540~kpc and its morphology is very peculiar, neither matching a head-tail AGN nor a shock-driven relic very well. For this reason, our study also includes polarimetric measurements of the A781 Main cluster. Rotation Measure analysis of the magnetic field depth is one way to probe the medium through which radio emission propagates. To this end we calibrated for the ellipticity of the telescope feeds and leakages stemming from the non-orthogonality of the feeds and synthesized images for the central cluster region using Briggs -0.5 weighting. We made a frequency cube at the native resolution of \textit{L}-band and {performed Rotation Measure (RM) synthesis} to recover the intrinsic polarization of the cluster members. We follow the definition and conventions defined by \cite{burn1966depolarization} and Faraday Depth synthesis derivation of \cite{brentjens2005faraday}. { The Full Width at Half Maximum (FWHM) of the Rotation Measure Transfer Function (RMTF)} is given for the Westerbork \textit{L}-band correlator\footnote{Digitized bandwidth coverage 1.301 to 1.460~GHz} as:

\begin{equation*}
\text{RMTF}_\text{FWHM}\approx \frac{2\pi}{\lambda^2_\text{max}} = 118.38~ \text{rad}~\text{m}^{-2},
\end{equation*} 

with a maximum function support for the channelizer given by

\begin{equation*}
\phi_\text{sup} \approx \frac{\sqrt{3}}{\min{(\Delta\lambda^2)}}= 4269.80~ \text{rad}~\text{m}^{-2}.
\end{equation*}
\vspace{1em}

We additionally deconvolve the Faraday Depth spectrum at each spatial pixel in the map using a variant of the \texttt{CLEAN} algorithm applied in Faraday Depth space to obtain the peak RM and peak-to-noise (PNR) on a pixel-by-pixel basis along the plane of the sky. This analysis was performed on the high-resolution 21~cm data due to the resolution limitations of the 92~cm data.

\begin{figure*}
\centering
\includegraphics[width=0.49\textwidth]{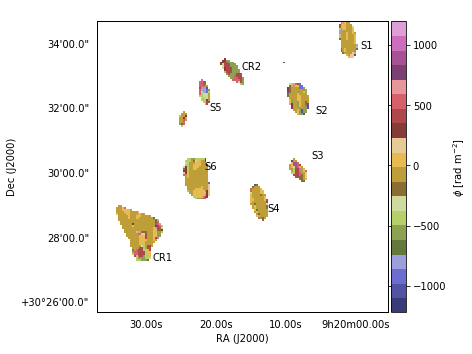}
\includegraphics[width=0.49\textwidth]{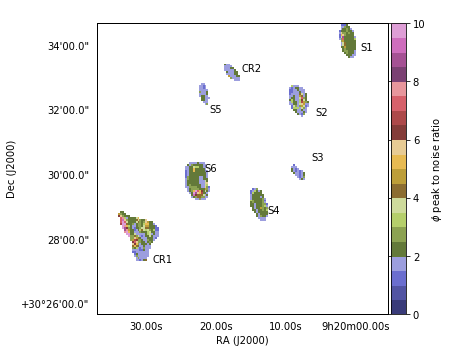}
\caption{{\textit{Left}: Peak rotation measure map of the Main cluster. \textit{Right}: Peak-to-noise map of the rotation measure values after RM deconvolution.}}
\label{fig:RM_map}
\end{figure*}

\begin{figure}
\centering
\includegraphics[width=0.45\textwidth]{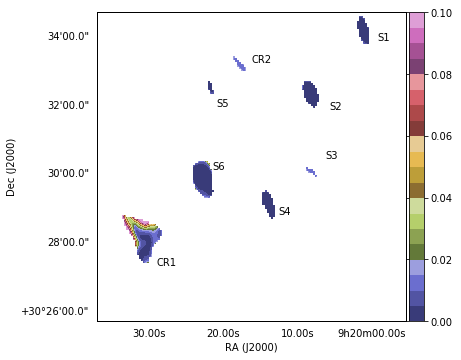}
\includegraphics[width=0.45\textwidth]{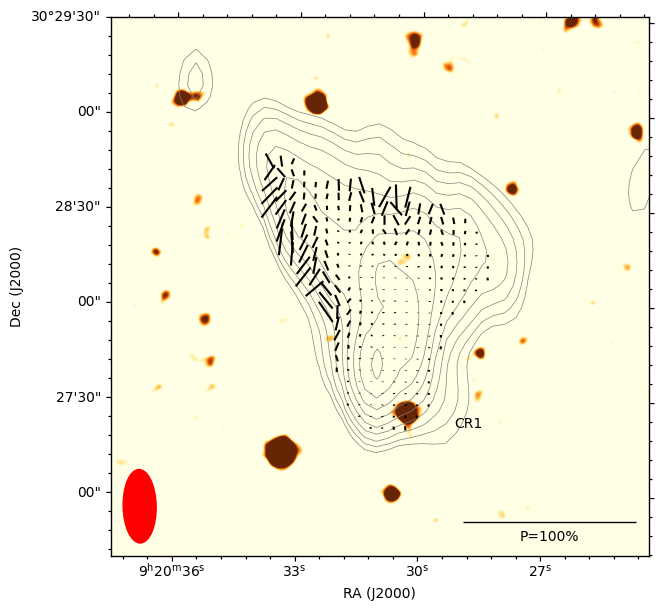}
\includegraphics[width=0.45\textwidth]{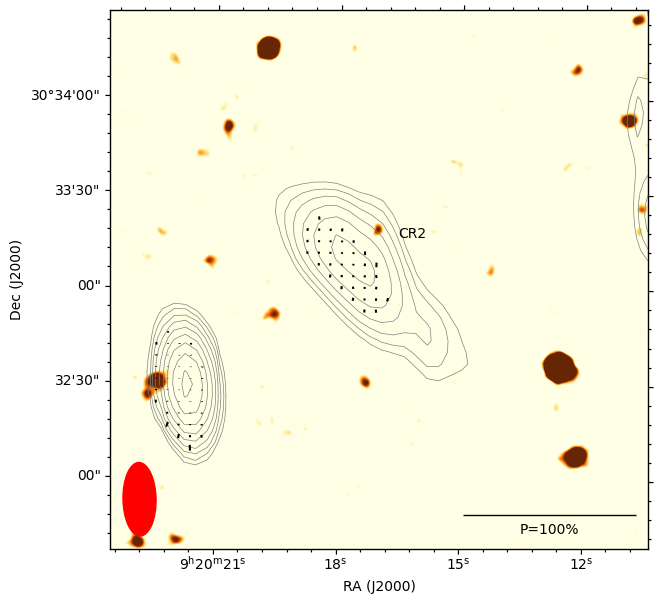}
\caption{\textit{Top}: polarization fraction map of the cluster. The cluster members are mostly depolarized apart of predominantly CR1 and to a lesser extent CR2 and S3. The colour bar indicates fractional polarization. Here 15$\sigma$ clips based on the total intensity map are used to differentiate the sources from the background. \textit{Middle} and \textit{Bottom}: zoom in to showcase the polarized edges of CR1 and CR2 respectively. Here the vector field is overlaid on the optical DSS red plate, with contours in steps of $\sqrt{2}$ from 13 sigma ($1\sigma=60~\mu$Jy~beam$^{-1}$)}
\label{fig:polplusCR1CR2}
\end{figure}

It is also important to note here {that the (linear) Electric Vector Polarization Angle (EVPA) calibration procedure discussed does not correct the ionospheric-induced RM on the EVPA, nor does it correct the absolute angle of the system receivers\footnote{See discussion in \citet{hales2017calibration}}}. Referring back to the widely-used assumption that 3C286 has a frequency constant EVPA of $33^\circ$ (with RM therefore very close to $0~\text{rad~m}^{-2}$) we estimate the ionospheric RM to be  $+4.610\pm0.922~\text{rad~m}^{-2}$. This gives a reasonably small offset at the centre of the narrow WSRT band of around $7^\circ$ from the assumed model. The angles and the quoted RM have been corrected for this contribution. The apparent recovered EVPA and fractional linear polarization are shown in Fig.~\ref{fig:polplusCR1CR2} for a cropped region around the cluster centre. We note that the linear polarization vectors shown for CR1 and CR2, the RM peaks in Fig.~\ref{fig:RM_map} and the associated statistics in Table~\ref{tab_rmsources} are corrected for both the approximate ionospheric contribution to the Faraday rotation, as well as the approximated $16 \pm 5$~rad~m$^{-2}$ Galactic foreground contribution \citep{oppermann2015estimating}.

The synthesized global RM map and associated peak-to-noise estimates are given in Fig.~\ref{fig:RM_map}. There is clear evidence for compressed polarized emission along the bright eastern spine and along the northwestern edge of the CR1 complex (see Fig.~\ref{fig:polplusCR1CR2}).

\begin{figure*}
\centering
\includegraphics[width=0.37\textwidth]{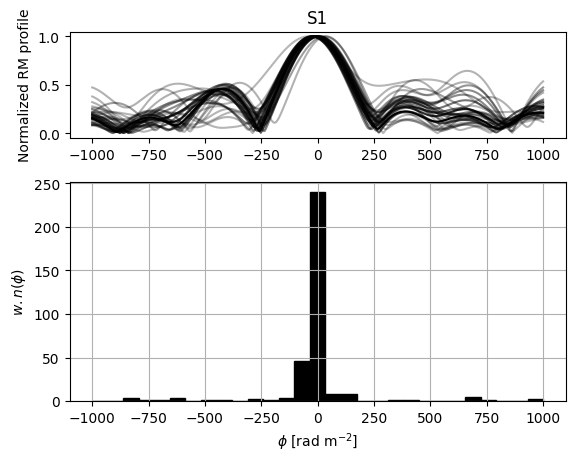}
\includegraphics[width=0.37\textwidth]{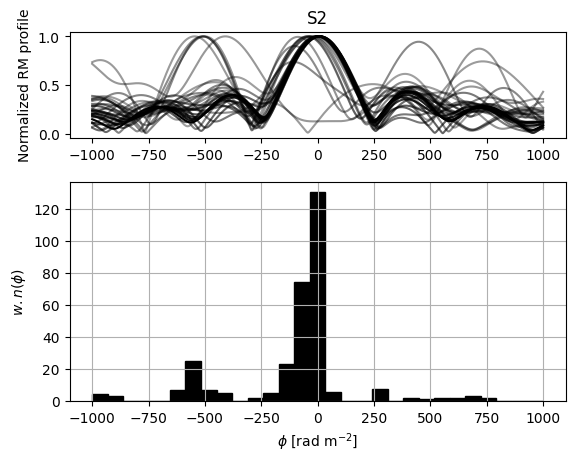}
\includegraphics[width=0.37\textwidth]{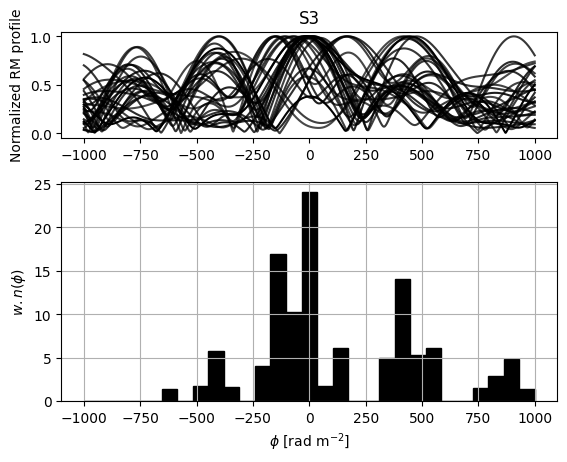}
\includegraphics[width=0.37\textwidth]{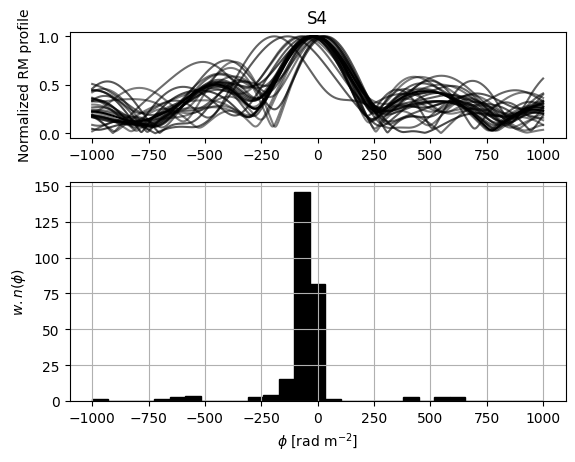}
\includegraphics[width=0.37\textwidth]{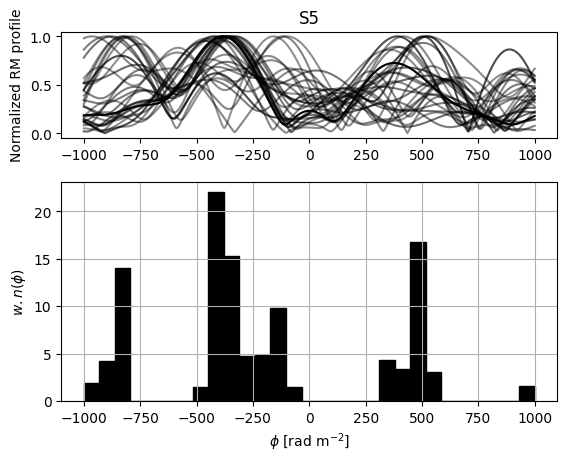}
\includegraphics[width=0.37\textwidth]{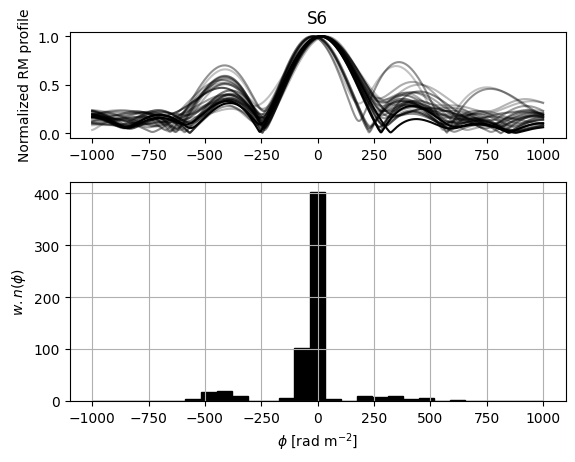}
\includegraphics[width=0.37\textwidth]{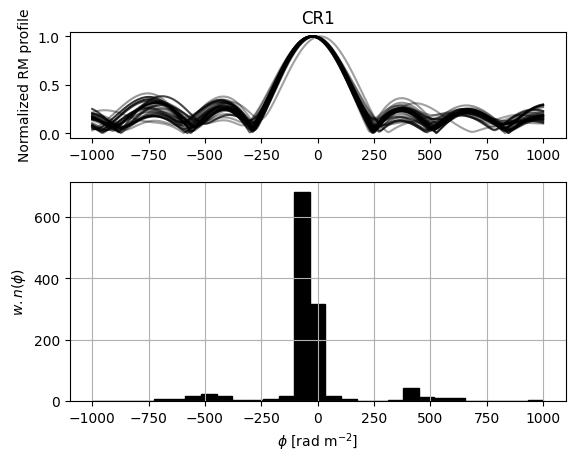}
\includegraphics[width=0.37\textwidth]{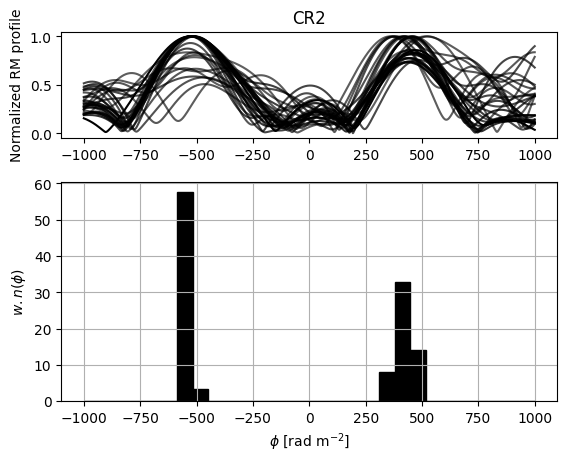}
\caption[RM spectra]{{\textit{Bottom}: Peak RM distributions (weighted by peak to noise) for a selection of sources, marked in the maps in fig.~\ref{fig:RM_map}. Clips are imposed to highlight only the AGN and candidate relic emission. We also plot the normalized non-deconvolved spectra for the 30 spectra with the highest deconvolution peak-to-noise ratio (PNR) (shown top right), extracted per source. The spectra' colours are dependent on the PNR weight --- darker to be interpreted as spectra with a higher PNR value. { SU1, S5 and S3 have low (as in the case of SU1) to moderate (S5 and S3) signal-to-noise ratios for RM measurement and the variations seen across the sources may not be an indication of substantial magnetic fields --- we do not plot spectra for these here.} The distributions consider pixels where the peak RM is at least 5x the noise, as indicated top right.}}
\label{fig:RM_map_spectra}
\end{figure*}

The compact sources at cluster redshift, S4 and S6, have a median peak Faraday depth along the line of sight in the range of a few 10s of rad~m$^{-2}$ (see Fig.~\ref{fig:RM_map}) and median peak distribution statistics in Table~\ref{tab_rmsources}. To varying degree, the integrated Faraday Depth spectra shown in Fig.~\ref{fig:RM_map_spectra} indicate that, apart from S4 and S6, the other established cluster members (S3 and S5) are not subject to a single constant Faraday screen. They instead show complex magnetic fields, both parallel and orthogonal to the line of sight. {The line of sight to the non-cluster-member S1 crosses the periphery to the cluster and by chance has a similar Faraday Depth to S4 and S6, both having lines of sight that may also only cross part of the whole cluster ICM.}

The majority of these sources are also largely depolarized, with the exception of CR1. This is not unexpected --- actively merging clusters tend to show little polarization within the merger region. This is likely due to the fine spatial-scale turbulence in such systems, where the synthesized beam acts to depolarize emission on these scales \citep{van2019diffuse}. In this case, the synthesized beam corresponds to about 46~x~105~kpc --- ie. similar in angular extent to most of the field sources.

Next, we will discuss the properties of these two candidate relics separately.

\begin{table}
  \centering
  \begin{tabular}{lllll}
  \hline \\
  Source & $z$ & $\phi$ max & $\phi$ median & $\phi$ IQS\\
  \hline \\
  S1 & 1.30526$^{\rm s}$ & -34.48 & -18.71 & 17.08\\
  S2 & - & -34.48 & -60.46 & 143.26 \\
  S3 & 0.30284$^{\rm s}$ & -34.48 & 0.26 & 541.27\\
  S4 & 0.29741$^{\rm s}$ & -103.45 & -43.38 & 34.16\\
  S5 & 0.30360$^{\rm s}$ & 448.28 & -339.40 & 868.12\\
  S6 & 0.29262$^{\rm s}$ & -34.48 & -20.61 & 24.67\\
  CR1 & - & -103.45 & -35.79 & 45.54\\
  CR2 & - & -586.21 & 394.95 & 948.77\\
  \hline
  \end{tabular}
  \caption[RM pixel distributions]{Properties of RM pixel distributions where rotation measure pixel peak to noise exceeds 3x, as shown in Fig.~\ref{fig:RM_map}. We indicate the peak of the distribution, as well as the median and IQS of the contributing data here.}
  \label{tab_rmsources}
\end{table}

\subsection{CR1 (head-tail galaxy or relic?)}
This source was already observed \citep{venturi2011elusive} at 325~MHz and tentatively classified as a candidate radio relic in the light of its peripheral position, morphology and spectral index steepening from $-1.4 < \alpha < -1.8$ northwards with decreasing distance from the cluster centre. \cite{botteon2019spectacular} identifies an optical counterpart to the west of the bright optical source near the peak of the radio emission seen in Fig.~\ref{fig:21_uniform_optical}. The source, spanning 550~kpc, is similar in its morphology when compared to observations taken at 150~MHz by \citet{botteon2019spectacular}. A bright knot of emission appears at the southernmost point of the CR1 source, connected to a high surface brightness spine that extends northeast. Two optical galaxies --- at approximately the cluster redshift --- coincide with the bright knot. Based on the combined X-ray and radio analysis, \cite{botteon2019spectacular} concluded that CR1 is either a relic or a head-tail radio galaxy with morphology distorted by a (weak) shock.

\begin{figure}
\centering
\includegraphics[width=.45\textwidth]{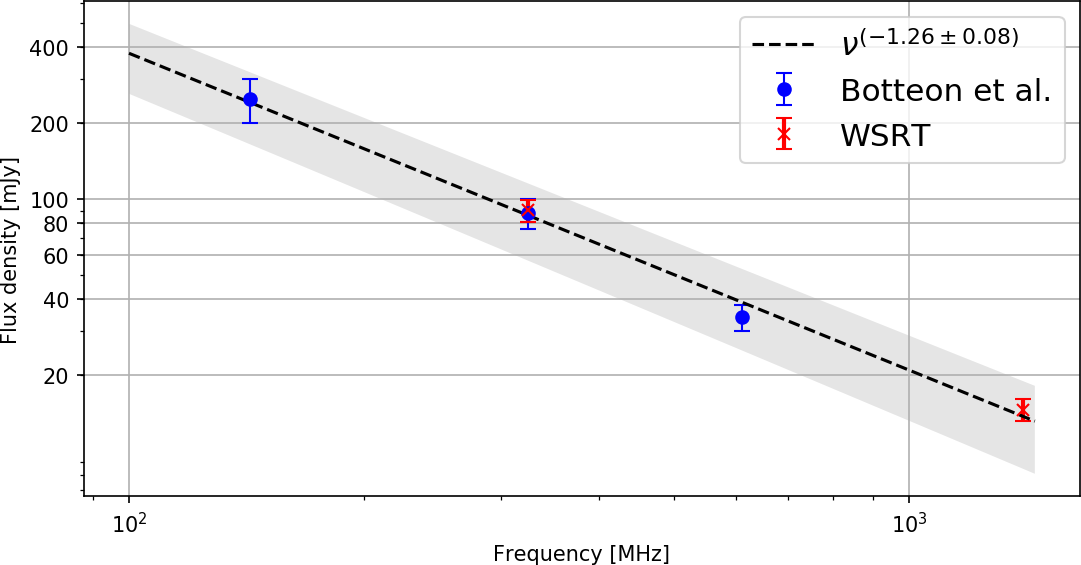}
\caption[Integrated spectrum of the source CR1]{Integrated spectrum of the source CR1. Here we show the measurements
made by \cite{botteon2019spectacular} and those we derive from the WSRT 92 and 21~cm bands. The dashed line indicates the best fitting to the LOFAR, uGMRT and WSRT flux density measurements to date. The shaded area is the propagated error on our fit. The new spectral index estimate of $-1.26 \pm 0.08$ is flatter than reported in \cite{botteon2019spectacular}}. 
\label{fig_CR1_power}
\end{figure}

Our total intensity image is in fair agreement with previous observations. The morphology of CR1 at 1.4~GHz is similar to the low-frequency data, with a similar extent ($\sim 540$~kpc linear size). We measure a flux density of $S^{CR1}_{1.4} = 14.6~\text{mJy}$, integrated over the the $60~\mu\text{Jy~beam}^{-1}$ contour shown in Fig.~\ref{fig:21_uniform_optical}. {The integrated flux density at 375 MHz ($90\pm9~\text{mJy}$) is in agreement with previous measurements ($88\pm12~\text{mJy}$) \citep{botteon2019spectacular}.} This is shown in Fig.~\ref{fig_CR1_power}.

\begin{figure*}
\centering
\includegraphics[width=0.45\textwidth]{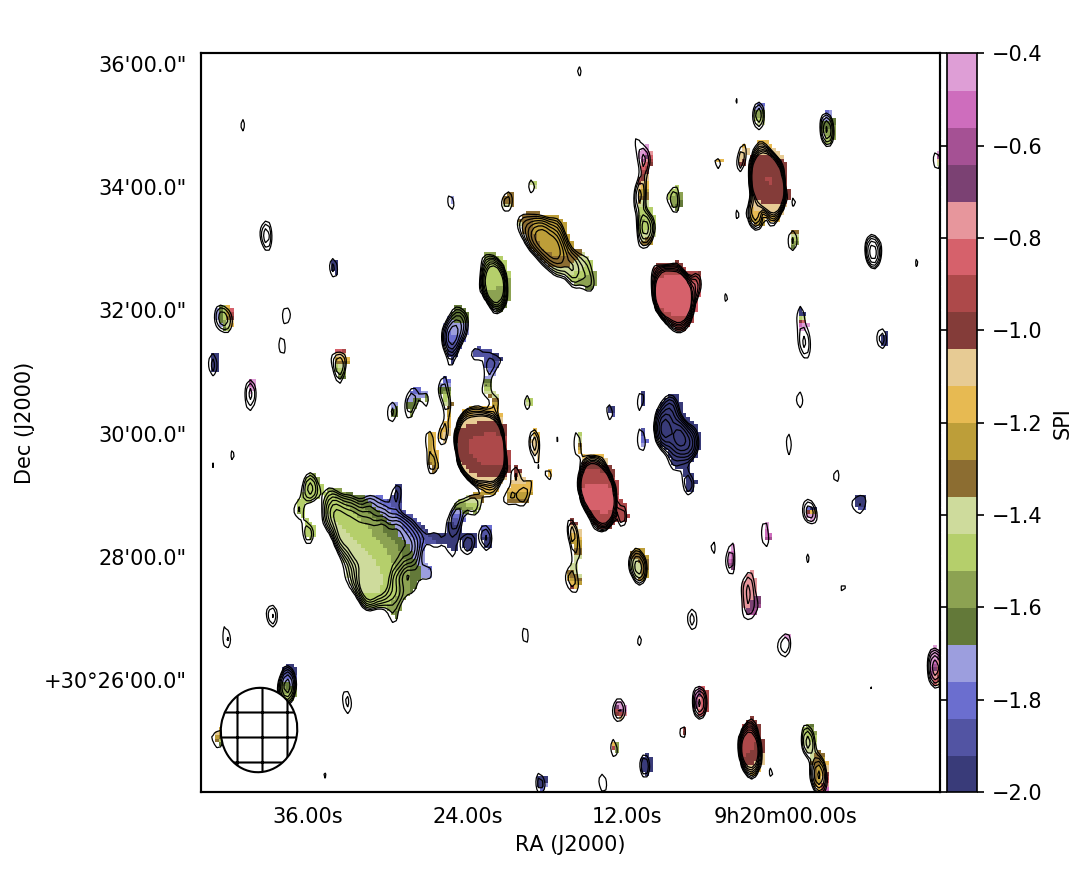}
\includegraphics[width=0.45\textwidth]{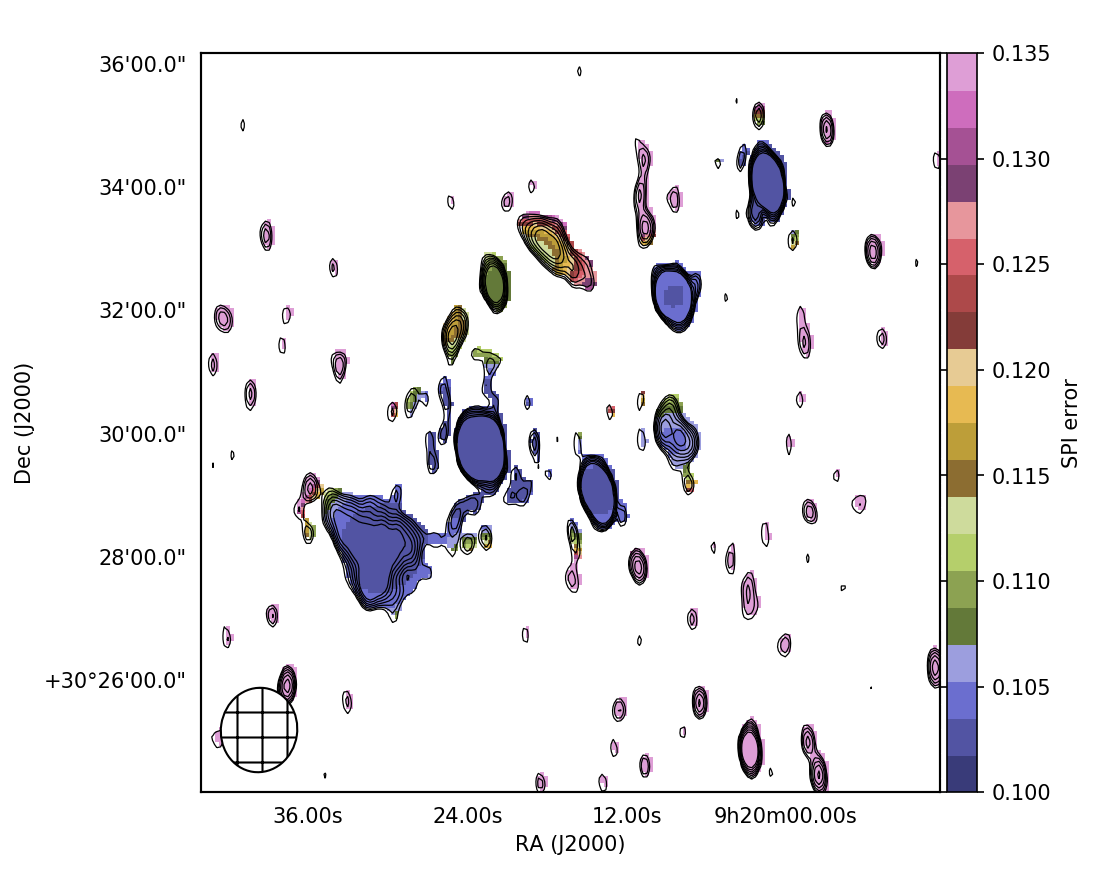}
\includegraphics[width=0.45\textwidth]{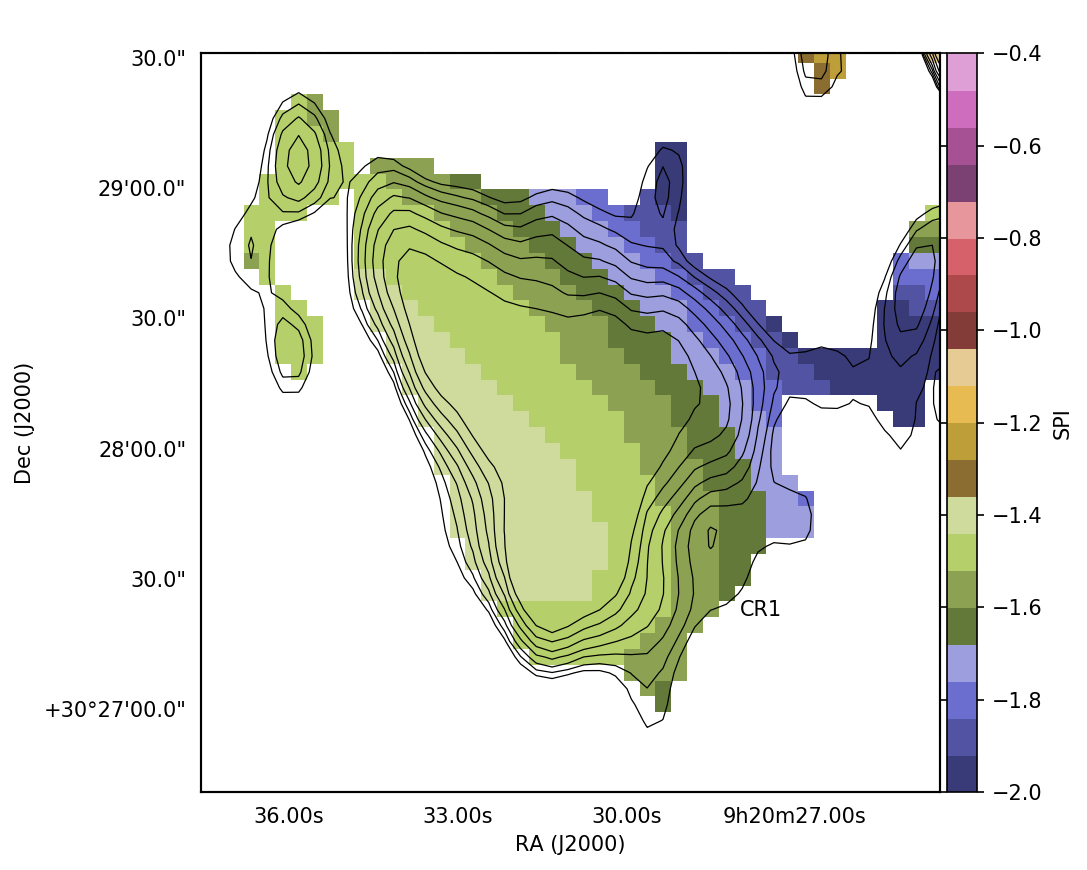}
\includegraphics[width=0.45\textwidth]{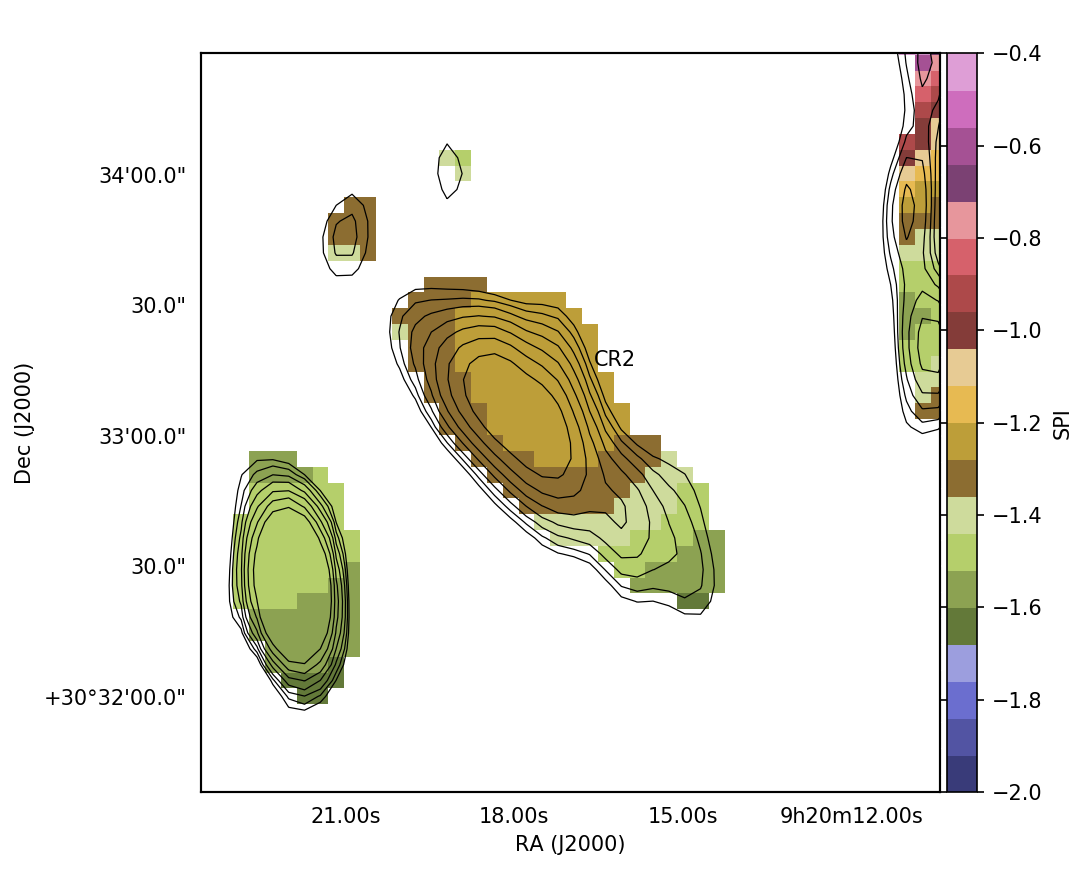}
\caption{{\textit{Top left:} Spectral index map computed from WSRT 92 and 21~cm maps after tapering the resolution to a circular Gaussian of $73.08''$ (lowest resolution) as indicated in the top map. The black contours from the uniform-weighted maps are drawn in steps of $\sqrt{2}$ starting at 60~$\mu$Jy~beam$^{-1}$. We apply
a dilated mask to the levels of the contours drawn to highlight only the spectral index in the region of the sources. \textit{Top right:} Associated spectral index error map. We assume standard quadrature propagation of error rules (with band co-variance assumed as 0) for logarithms with flux scale errors at 10\% level and tapered 21- and 92-cm noise of 18~$\mu$Jy~beam$^{-1}$ and 1~mJy~beam$^{-1}$ respectively. \textit{Bottom left} and \textit{Bottom right}: Zoom in of the CR1 and CR2 complexes respectively.}}
\label{fig:SPI}
\end{figure*}

Assuming the source is within the vicinity of the Main cluster at redshift $z=0.3004$ the power extrapolation is slightly underestimated at 1.4~GHz by \cite{botteon2019spectacular} (who assumed a spectral index steeper than presented here). The spectrum is plotted in Fig.~\ref{fig_CR1_power} --- here we use the integrated spectrum of the bright emission of the  spine, which is about -1.4 and the integrated emission within the first contour of the CR1 complex in Fig.~\ref{fig:21_uniform_optical} to derive the radio power at 1.4GHz:

\begin{equation*}
    P_{\text{CR1},1.4\text{GHz}} = \frac{4\pi D_\text{L}^2 S_\nu}{(1+z)^{1+\alpha}} \approx 4.78\times10^{24}~\text{W}~\text{Hz}^{-1}.     
\end{equation*}
The spectral index image is shown in Fig.~\ref{fig:SPI}. We used the same image field of view and sampling for both maps and tapered both maps to the lowest possible resolution, as taken from the 92~cm fitted beam (83~arcsec). We also corrected for the attenuation of the antenna primary lobe, using Equation~\ref{eqn:wsrtpbeam}. A 4~mJy~beam$^{-1}$ cutoff was used to allow for only high sigma components in the spectral index image.

Although the 92~cm resolution does not resolve the fine structure of either candidate (CR1 or CR2) the integrated trends are visible. As noted in \cite{botteon2019spectacular} the bulk of the spine of CR1 has a steep spectrum of $\alpha \approx -1.4$, steepening to $\alpha<-1.6$ closer to the cluster centre. This is indicative of synchrotron radiation losses. The moderately steep spectra are consistent with the known spectra of other relics \citep{van2019diffuse}.

From the RM map (Fig.~\ref{fig:RM_map}) we see that the area near the overlapping optical galaxy immediately northwest of the bright radio bulge to the south is largely depolarized (Fig.~\ref{fig:polplusCR1CR2}). On the contrary, the rest of the CR1 complex is relatively polarized (Fig.~\ref{fig:polplusCR1CR2}), especially the eastern and western edges. Both edges have peak rotation measures closer to zero and agree with what is observed on other compact sources at the cluster redshift, specifically S6 and S4.

The polarization EVPA is reasonably well aligned in the plane of the sky in the direction we would expect to see a merger shock (Fig.~\ref{fig:polplusCR1CR2}). However, the increased polarization fraction along the structure is more consistent with relatively low fractional polarization as is typically observed in the jets of AGN \citep{homan2005polarization}. The polarization characteristics of this peripheral complex stand in stark contrast to the degree of polarization of typical relics observed in the literature (see e.g. \cite{wittor2019polarization}).

The spectral index  is substantially steeper than expected for steep spectra AGN population at 1.4~GHz (e.g. \cite{de2010radio}) --- this source will be considered very steep according to the distributions of SPI for both narrow and wide-tailed radio AGN \citep{sasmal2022new}, while the spectrum is at the low end of expected integrated spectra for radio relics \cite{feretti2012clusters}.  Although the low polarization fraction observed and the physical size (assuming cluster redshift for this source) may point to the source belonging to the class of \textit{Radio Pheonix} (shock reaccelerated fossil emission from AGN), the ultra-steep spectra of these sources are typically curved and in excess of -1.5 \cite{van2019diffuse}. As a result, such emission is typically only seen at much longer wavelengths. Coupled with the coincidence of the optical counterpart with the knot of radio emission in the south of the complex, as reported by \cite{botteon2019spectacular}, the spectral and polarimetric measurement suggests that CR1 is neither a radio relic nor Pheonix. It is much more likely that CR1 is an ageing head-tail galaxy.

\subsection{CR2: another candidate relic?}
There is extended emission north of the hot X-ray region, labelled as CR2. This is the same source seen by \cite{botteon2019spectacular} in LOFAR data. The elongated source is roughly $1.5'$ ($\sim 405$~kpc, assuming cluster redshift), with an integrated flux density, of $S_{\rm CR2, 21cm} = 2.7\pm0.27$~mJy, measured within the area defined by the $60~\mu\text{Jy}~\text{beam}^{-1}$ contour in Fig.~\ref{fig:21_uniform_optical}. The source has a spectrum between $-1.0$ to $-1.2$ over most of its area, as seen in Fig.~\ref{fig:SPI} and there is no clear optical counterpart (Fig.~\ref{fig:polplusCR1CR2} Bottom). Similarly to CR1, the radio spectrum of CR2 steadily steepens towards the the cluster center.

If we assume an integrated spectral index of $-1.1$, and that the source has the same redshift as the cluster, then the k-corrected radio power is

\begin{equation*}
    P_{\text{CR2},1.4\text{GHz}} \approx 8.16\times10^{23}~\text{W}~\text{Hz}^{-1}. 
\end{equation*}
CR2 is only slightly polarized (Fig.~\ref{fig:polplusCR1CR2} bottom) and has a rotation measure markedly different to the primary cluster AGNs S3, S4 and S6 (Fig.~\ref{fig:RM_map_spectra}). Considering the contrast of the distribution of Faraday Depths of CR2 compared to the other cluster sources, save for S3 and S5, it is clear that this source of emission must either be located in an area with marked differences in foreground magnetic fields or intrinsically has complex Faraday screens. However, both its compact round morphology and low fractional polarization is in stark contrast to what is generally expected for radio relics, although we cannot exclude projection effects on the morphology due to the available resolution.

\subsection{Revisiting halo claims}
\cite{botteon2019spectacular} achieves sensitivities of $\sigma_\text{143MHz}=270~\mu\text{Jy}~\text{beam}^{-1}$, $\sigma_\text{325MHz}=150~\mu\text{Jy}~\text{beam}^{-1}$,
$\sigma_\text{610MHz}=120\mu\text{Jy}~\text{beam}^{-1}$ at resolutions of  $11.1\times6.5~\text{arcsec}$, $10.6\times7.2~\text{arcsec}$ and $13.5\times 9.8~\text{arcsec}$ respectively. At $12~\mu\text{Jy}~\text{beam}^{-1}$ rms noise at a resolution of $23.2\times10.4~\text{arcsec}$, using all redundant spacings, our 21~cm data is the most sensitive \textit{L}-band data on this field to our knowledge, and is at similar magnitude to LOFAR sensitivities when scaled by a halo spectrum of -1.3. This improved \textit{L}-band sensitivity warrants a renewed look at the cluster center for any signs of halo emission.

\begin{figure}
\centering
\includegraphics[width=0.49\textwidth]{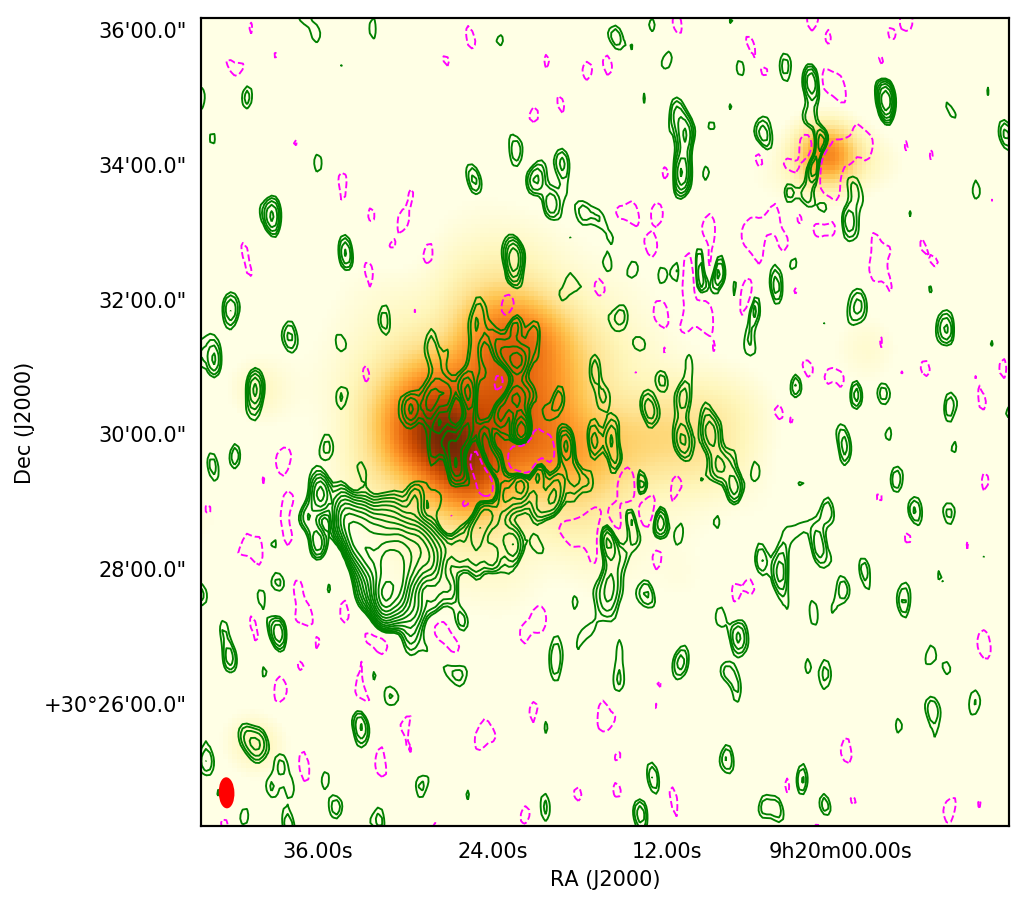}
\includegraphics[width=0.49\textwidth]{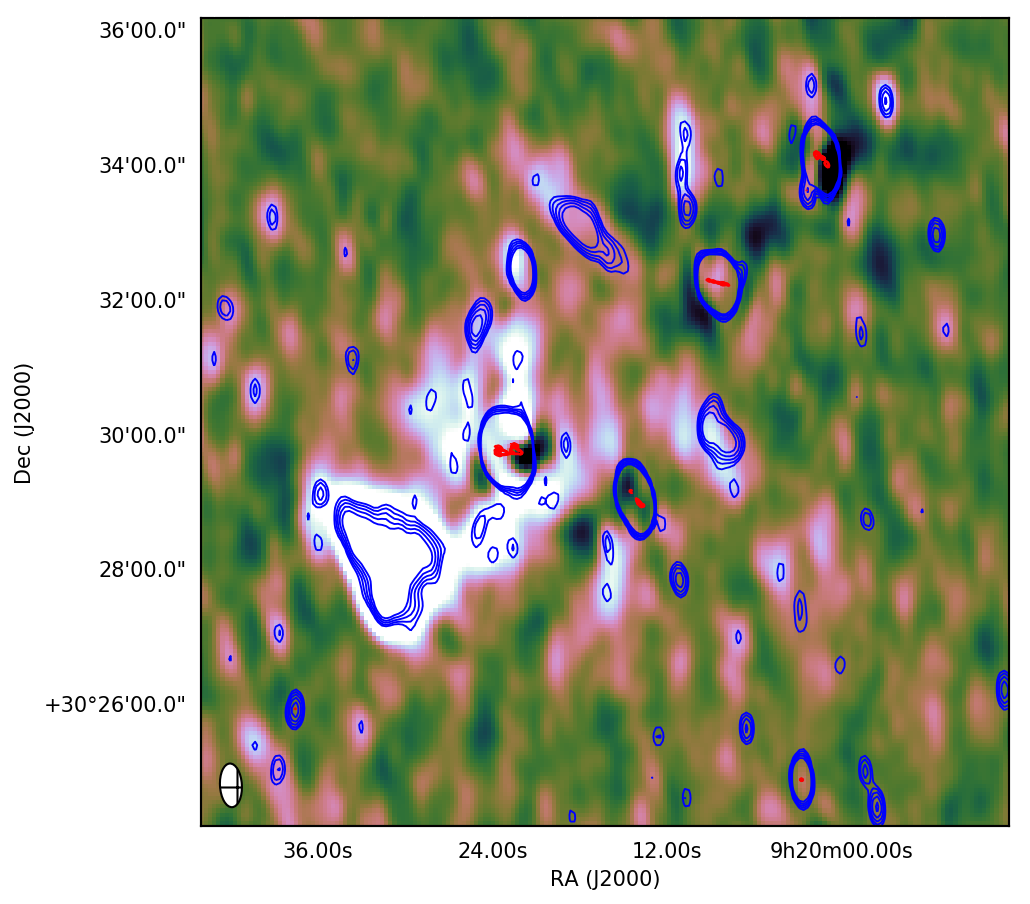}
\caption[subtracted overlays]{{\textit{Top:} Briggs -0.25 weighted residuals after subtracting the bright AGNs in the cluster overlaid on \textit{XMM-Newton} X-ray. The rms noise is estimated as 6.23~$\mu$Jy~beam$^{-1}$. Radio contours are plotted in $\sqrt{2}$-scale starting at $6~\sigma$. Dashed contours indicate $-6~\sigma$ subtraction errors. The synthesized beam at -0.25 weighting is $25.8''\times 12.1''$ ($\Omega_b\approx26.63\text{px}$) shown in red.
\textit{Bottom:} Subtracted residuals plotted in colour overlaid with blue contours from
uniform maps presented in Fig.~\ref{fig:final_21_image} for reference, contours starting from $7~\sigma$ ($\sigma = 12~\mu\text{Jy}~\text{beam}^{-1}$) in $\sqrt{2}$-scale. Red contours from the higher resolution VLA reductions overlaid, starting from $7~\sigma$ ($\sigma \approx 60~\mu\text{Jy}~\text{beam}^{-1}$) in $\sqrt{2}$-scale. Beam indicated in hatched lines.}}
\label{fig:21_subtracted_xray}
\end{figure}

Fig.~\ref{fig:21_subtracted_xray} shows the 21~cm Briggs $-0.25$ weighted residuals after the bright AGNs within the cluster have been subtracted from the visibilities by means of a {Direct Fourier Transform (DFT)} implemented with \textsc{Meqtrees} \citep{noordam2010meqtrees}. The subtraction is performed by iterative fitting for components above $2\sigma$ using \textsc{pyBDSF} \citep{mohan2015pybdsf} and re-imaging. In total 3 rounds of subtraction and re-imaging were performed on the 21~cm data, each time explicitly excluding the components fitted to the southeastern complex. We carefully checked that the subtracted components only fall within the areas of AGNs S2--S6, to within instrument resolution. All fitted components were delta components. The accuracy in subtracting S6 ($S_{1.4} \sim 32.6\pm3.3$~mJy~beam$^{-1}$ peak flux density) is the limiting factor in achieving high dynamic range on the low-resolution images {within the vicinity of the hot merger region.}

{ To the south of the Main cluster  we find traces of a bridge-like structure connecting CR1 to low-level emission extending throughout the hot X-ray plasma surrounding the central AGN}. Excluding this bridge-like low-level structure we find the integrated flux of the extended emission to be $1.9\pm0.2~\text{mJy}$ within the $6\sigma$ contours {in the X-ray-radio plot in Fig.~\ref{fig:21_subtracted_xray} (top). This measurement is limited by a subtraction error at the level of 0.1~mJy~beam$^{-1}$. It should be cautioned that the instrumental resolution of the WSRT system is the main limiting factor to the case presented here and spurious background sources, coupled with subtraction errors from S6 may be contributing to the integrated flux.}

{If the apparently-diffuse emission seen is indeed a halo, its integrated flux would place it an order of magnitude below the upper bounds set by \cite{venturi2011elusive} and \cite{botteon2019spectacular}.} It is in-keeping with an average spectral index below $-1.44$ if the upper detection limit of $S^\text{upper}_\text{143MHz} = 50~\text{mJy}$ is assumed from \cite{botteon2019spectacular}. Assuming this as an upper limit to the spectral index, the k-corrected radio power at cluster redshift is:

\begin{equation*}
    P_{\text{Halo},1.4\text{GHz}} \approx 6.28\times 10^{23}~\text{W}~\text{Hz}^{-1}.
\end{equation*}
Although one would have expected this power to be at least an order of magnitude larger on the $0.1-2.4$~kev X-ray / radio power correlation, the power is still close to other detected haloes on the Mass / radio power correlation \citep{van2019diffuse}. {Without the availability of sensitive higher resolution data the error estimates presented here may be optimistic, however, it is noted that the apparently-diffuse emission extends over the majority of the disturbed X-ray thermal region with a diameter (excluding the bridge-like structure) of around 0.6 Mpc.}

\section{Conclusions}
{We have observed the dynamically-disturbed A781 cluster complex with the Westerbork Synthesis Radio Telescope at 21 and 92 cm. We presented the most sensitive \textit{L}-band observations of the system to date.}

{ We have found, what appears to be the existence of low-level diffuse emission around the central region of the merging cluster, although our measurement is limited by instrumental resolution. The integrated emission is nearly an order of magnitude less than the flux density claimed by \cite{govoni2011large}.} This is in-keeping with \citep{venturi2011observational} of an unusual flat-spectrum radio halo and is well below the expected radio power predicted by the $P_\text{1.4GHz} - L_x$ relationship. 

Our maps corroborate the \cite{botteon2019spectacular} observation of radio emission at the southeastern and northwestern flanks of the hot X-ray plasma. We have studied the polarimetric properties of the southeastern and northern complexes in detail. We find that the edges of the southeastern complex are polarized, with low Faraday Depth. Neither complex is highly polarized (fractions less than 8\% and 1.5\% for the southern and northern complexes respectively), further qualifying earlier statements by \cite{botteon2019spectacular} that only relatively weak shocks are present in the Main cluster. This evidence, including consideration of morphology, points to the contrary that these are radio relics. The southeastern complex most likely has its origin as head-tail emission from an AGN with an unclear optical counterpart.

{ The corroborating evidence hinting to the existance of an ultra-low flux density halo warrants further telescope time with sensitive high-resolution instruments such as LOFAR at lower frequencies, and SKA precursor telescopes such as the MeerKAT \textit{UHF} (544--1088~MHz) and \textit{L}-band (856--1712~MHz) systems. Such observations will firmly establish the spectrum and integrated power of this very peculiar cluster.}

\section*{Data availability}
Data was generated at a large-scale facility, WSRT. FITS files are available from the authors upon request.

\section*{Acknowledgements}
This work is made possible by use of the Westerbork Synthesis Radio Telescope operated by ASTRON Netherlands Institute for Radio Astronomy.
Our research is supported by the National Research Foundation of South Africa under grant 92725. Any opinion, finding and conclusion or recommendation expressed in this material is that of the author(s) and the NRF does not accept any liability in this regard. This work is based on the research supported in part by the National Research Foundation of South Africa (grant No. 103424).
This research has made use of the services of the ESO Science Archive Facility.
The Second Palomar Observatory Sky Survey (POSS-II) was made by the California Institute of Technology with funds from the National Science Foundation, the National Geographic Society, the Sloan Foundation, the Samuel Oschin Foundation, and the Eastman Kodak Corporation. 
Based on observations obtained with \textit{XMM-Newton}, an ESA science mission with instruments and contributions directly funded by ESA Member States and NASA.
This research has made use of the VizieR catalogue access tool, CDS, Strasbourg, France. The original description of the \textsc{VizieR} service was published in A\&AS 143, 23.
This research made use of \textsc{APLpy}, an open-source plotting package for \textsc{Python} hosted at http://aplpy.github.com
This research made use of \textsc{Astropy}, a community-developed core \textsc{Python} package for Astronomy (Astropy Collaboration, 2013). The research of OS is supported by the South African Research Chairs Initiative of the Department of Science and Technology and National Research Foundation.
The primary author wishes to thank the National Research Foundation of South Africa for time granted towards this study.
\bibliographystyle{mnras}
\bibliography{a781} 

\bsp	
\label{lastpage}
\end{document}